\documentclass[conference,keeplastbox]{IEEEtran}
\usepackage{cite}
\usepackage{amsmath,amssymb,amsfonts}
\usepackage{algorithmic}
\usepackage{graphicx}
\usepackage{textcomp}
\usepackage{xcolor}
\usepackage{float}
\usepackage{subfigure}
\PassOptionsToPackage{hyphens}{url}
\usepackage{hyperref}

\usepackage{graphicx}  

\usepackage{caption}

\DeclareCaptionLabelFormat{lc}{{#1}~#2}
\makeatletter
\def\endthebibliography{%
  \def\@noitemerr{\@latex@warning{Empty `thebibliography' environment}}%
  \endlist
}
\makeatother

\usepackage{epsfig,endnotes}

\usepackage{floatpag,enumitem}

\usepackage{relsize}

\usepackage{tikz}

\usepackage{multirow}
\usepackage{setspace}
\usepackage{mathrsfs}
\usepackage{bbm}
\usepackage{pifont}

\usepackage{amsfonts}

\usepackage{amsmath, amsthm}

\usepackage{tabularx}
\usepackage{listings}
\usepackage{graphicx}
\usepackage{amssymb,booktabs}
\usepackage{array}
\usepackage{cases}

 \usepackage{supertabular}
\usepackage{mathrsfs}

\usepackage{psfrag}
\usepackage{bbding}

\usepackage{float}

\usepackage{amsbsy}

\makeatletter
\providecommand{\leftsquigarrow}{%
  \mathrel{\mathpalette\reflect@squig\relax}%
}
\newcommand{\reflect@squig}[2]{%
  \reflectbox{$\m@th#1\rightsquigarrow$}
}
\makeatother

\usepackage{algorithm}
 \usepackage{algorithmic}

\makeatletter
\newcommand{\newalgname}[1]{%
  \renewcommand{\ALG@name}{#1}%
}

\newcommand {\C} {{\rm I\kern-5.5pt C}}

\def\centerhack#1{\hbox to 0pt{\hss\footnotesize #1\hss}}
\def\centerhackn#1{\hbox to 0pt{\hss #1\hss}}
\def\dchack#1{\vbox to 0pt{\vss{\hbox to 0pt{\hss#1\hss}}\vss}}

\usepackage{etoolbox}
\AtBeginEnvironment{align}{\setcounter{subeqn}{0}}
\newcounter{subeqn} %

\usetikzlibrary{positioning}

\newcounter{mysub}

\newtheorem{lem}{Lemma}

\newtheorem*{proposition1.1}{Proposition 1.1}
\newtheorem*{proposition1.2}{Proposition 1.2}
\newtheorem*{proposition1.3}{Proposition 1.3}
\newtheorem*{proposition2.1}{Proposition 2.1}
\newtheorem*{proposition2.2}{Proposition 2.2}
\usepackage{subfigure}

\begin{document}

%

\title{Intelligent Reflecting Surface Meets Mobile Edge Computing: Enhancing Wireless Communications for Computation Offloading}
%
%
%
\author{\textup{Yang Liu}$^{1}$, \textup{Jun Zhao}$^{1}$, \textup{Zehui Xiong}$^{1}$, \textup{Dusit Niyato}$^{1}$, \textup{Chau Yuen}$^{2}$, \textup{Cunhua Pan}$^{3}$, \textup{Binbin Huang}$^{4}$\\
$\{$yang-liu,junzhao,zxiong002,dniyato$\}$@ntu.edu.sg, yuenchau@sutd.edu.sg, c.pan@qmul.ac.uk, huangbinbin@hdu.edu.cn\\
\IEEEauthorblockA{1: Nanyang Technological University\\
2: Singapore University of Technology and Design \\
3: Queen Mary University of London \\
4: Hangzhou Dianzi University
}}


%
%

\markboth{Journal of \LaTeX\ Class Files,~Vol.~14, No.~8, August~2015}%
{Shell \MakeLowercase{\textit{et al.}}: Bare Demo of IEEEtran.cls for IEEE Journals}
%



\maketitle

\begin{abstract}
We consider computation offloading for edge computing in a wireless network equipped with intelligent reflecting surfaces (IRSs). IRS is an emerging technology and has recently received great attention since they can improve the wireless propagation environment in a configurable manner and enhance the connections between mobile devices (MDs) and access points (APs).  At this point not many papers consider edge computing in the novel context of wireless communications aided by IRS. In our studied setting, each MD offloads computation tasks to the edge server located at the AP to reduce the associated comprehensive cost, which is a weighted sum of time and energy. The edge server adjusts the IRS to maximize its earning while maintaining MDs' incentives for offloading and guaranteeing each MD a customized information rate. This problem can be formulated into a difficult optimization problem, which has a sum-of-ratio objective function as well as a bunch of nonconvex constraints. To solve this problem, we first develop an iterative evaluation procedure to identify the feasibility of the problem when confronting an arbitrary set of information rate requirement. This method serves as a sufficient condition for the problem being feasible and provides a feasible solution. Based on that we develop an algorithm to optimize the objective function. Our numerical results show that the presence of IRS enables the AP to guarantee higher information rate to all MDs and at the same time improve the earning of the edge server.
\end{abstract}


\section{Introduction}

\textbf{Edge computing}. As an emerging computing paradigm, \textit{edge computing}~\cite{satyanarayanan2017emergence,mao2017survey} pushes computing tasks
to the network edges such as base stations and access points. Both edge computing and cloud computing can support computation-intensive applications. Yet, since edge servers are closer to mobile devices than clouds, edge computing is more suitable to enable latency-critical applications
at mobile devices~\cite{hu2018wireless}. Many companies are now leveraging edge computing to ensure that their services are available at the edges with high speed~\cite{vaughan2019working}. In July 2019,  AT\&T signed  a deal of US\$2 billion with Microsoft to use the latter's capabilities related to edge and cloud computing~\cite{Microsoft}.

\textbf{Computation offloading for edge computing.}
An important topic studied in edge computing is \textit{computation offloading}~\cite{mach2017mobile}, where a mobile device offloads intensive computation tasks to the edge server with stronger computational resources. In the seminal model for computation offloading proposed by~\cite{chen2014decentralized,chen2015efficient} by Chen and his co-authors, mobile devices choose computation offloading to the edge cloud or local computing on the devices by comparing the utilities under the two settings, where the utility is computed as a weighted function of the time and energy required to complete the computation task. This model, which our paper will use, has been adopted (with possible refinements) in many studies~\cite{tao2017performance,mao2016dynamic}.




\textbf{Intelligent reflecting surfaces (IRSs) and IRS-aided communications}. An \textit{intelligent reflecting surface (IRS)} can intelligently control the wireless environment to improve signal strength received at the destination. This is vastly different from prior techniques which improve wireless communications via optimizations at the sender or receiver. Specifically, an IRS consists of many IRS units, each of which can  reflect the incident signal at a reconfigurable angle. In such \textit{IRS-aided communications}, the wireless signal travels from the source to the IRS, is optimized at the IRS, and then travels from the IRS to the destination. Such communication method is particularly useful when the source and destination have a weak  wireless channel in between  due to obstacles or poor environmental conditions, or they do not have direct line of sights.

Because of the capability of configuring wireless environments, IRSs are envisioned by many experts in wireless communications to play an important role in 6G networks. In November 2018, the Japanese mobile operator NTT DoCoMo and a startup MetaWave demonstrated the use of IRS-like technology for assisting wireless communications in 28GHz band~\cite{DOCOMO}. IRSs have been compared with the massive MIMO technology used in 5G communications. IRSs reflect wireless signals and hence consume little power, whereas massive MIMO transmits signals and needs much more power~\cite{hu2018beyond}.





\begin{figure}[!t]
 \centering
\includegraphics[scale=0.35]{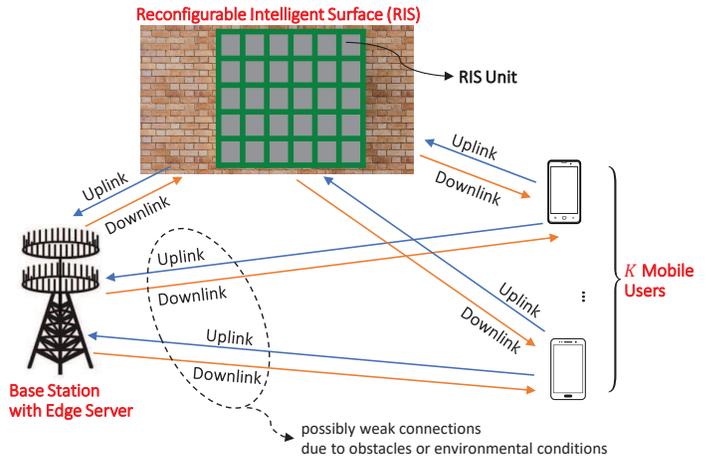}
\caption{An Illustration of IRS-Aided Communication System} \label{fig-system}
\end{figure}

\textbf{Problem studied in this paper: Computation offloading for
edge computing in IRS-aided communications.} We tackle the problem of mobile devices offloading computation tasks to a base station equipped with an edge server,
in the context of IRS-aided wireless communications. In the studied setting, mobile devices intend to send computation tasks to the base station via wireless channels, and communications between them are assisted by an IRS.
Fig.~\ref{fig-system} provides an illustration of the system model.

\textbf{Contributions.} The contributions of this paper are summarized as follows:
\begin{enumerate}

\item To the best of our knowledge, there exists no paper considering the computation offloading problem in the presence of IRS except that the very recent paper \cite{CunhuaPan_MEC_IRS} by Bai \emph{et al.} considers the latency minimization problem with the aid of IRS in edge computing system. In this paper, we extend the cost metric of mobile computing to a more comprehensive consideration, which subsumes the one used in \cite{CunhuaPan_MEC_IRS} as a special case. Besides we also take into account the rate constraints, which makes our problem more meaningful and much more challenging.  

\item To solve the problem, We first develop an algorithm to detect the feasibility when given an arbitrary set of information rate constraints. The algorithm can also provide a feasible solution when we identify the problem as feasible. Then we develop an efficient solution which can maximize the earning of the edge server.  

\item Through substantial numerical results, we demonstrate the significance of IRS devices. Our experimental results show that IRS can significantly boost the feasibility probability for information rate constraints. Besides, numerical results suggest that our algorithm converges fast and can effectively improve the edge server's earning.

\end{enumerate}


\textbf{Notation.} Scalars are denoted by italic letters, while vectors and matrices are denoted by bold-face
lower-case and upper-case letters, respectively. $\mathbb{C}$ denotes the set of all complex numbers. For a vector $\boldsymbol{x}$, $\|\boldsymbol{x}\|$ denotes its Euclidean norm. For a matrix $\boldsymbol{M}$, its transpose and conjugate transpose are denoted by $\boldsymbol{M}^T$ and $\boldsymbol{M}^H$, while $\boldsymbol{M}_{i,j}$ means the element in the $i$-th row and $j$-th column of $\boldsymbol{M}$. For a vector $\boldsymbol{x}$, its transpose, conjugate transpose, and Euclidean norm are denoted by $\boldsymbol{x}^T$, $\boldsymbol{x}^H$, and $\|\boldsymbol{x}\|$, while $x_{i}$ means the $i$-th element of $\boldsymbol{x}$.







\section{Related Work} \label{sec-Related-Work}


Since our paper combines edge computing and wireless communications aided by intelligent reflecting surfaces (IRSs), we discuss both related studies in edge computing and IRS-aided communications.



\textbf{Edge computing and computation offloading}.
Edge computing pushes computations to the network edges in order to support  latency-critical applications at mobile devices~\cite{hu2015mobile}. In many studies~\cite{mao2017survey}, edge computing is also referred to as \textit{fog computing}, a term introduced by Bonomi~\textit{et~al.}~\cite{bonomi2012fog} of Cisco. In 2018, IEEE integrated the fog computing reference architecture developed by the \textit{OpenFog Consortium} (now the \textit{Industrial Internet Consortium}) as a standard~\cite{IEEEOpenFog}.




Among many references on computation offloading for edge computing, two studies most related to our work are the seminal work~\cite{chen2014decentralized,chen2015efficient} by Chen and his co-authors. Chen~\cite{chen2014decentralized} considers a network of many mobile device users and one base station equipped with edge cloud. Mobile devices select computation offloading to edge cloud or local computing on the devices by comparing the obtained utilities under the two settings. All communications in the network use the same wireless channel and interfere each other. Hence, the data uploading rate of a mobile device selecting computation offloading is related to the offloading decisions of other devices. Then the dynamics in offloading decision making of mobile devices are modeled as a game, which is analyzed using potential game theory. Improving~\cite{chen2014decentralized} to a more general setting, Chen~\textit{et~al.}~\cite{chen2015efficient} address a
wireless interference environment of multiple channels instead of only one channel. Thus, the data uploading rate of a mobile device selecting computation offloading is related to other devices which use the same wireless channel. Then a game is used to decide local/edge computing choices of mobile devices and their channel selection in the case of edge computing.

In addition to~\cite{chen2014decentralized,chen2015efficient} discussed above, many studies on computation  offloading  for  edge  computing have appeared recently. We discuss some of them below and refer readers to~\cite{tao2017performance,wang2019edge} for more comprehensive reviews. Mao~\textit{et~al.}~\cite{mao2016dynamic} consider computation offloading of an energy-harvesting device to the edge server, where the incorporation of energy harvesting makes the design of computation offloading policy challenging. Partial computation offloading instead of  full offloading is investigated in~\cite{wang2016mobile,ning2018cooperative,tao2017performance,bi2018computation}. Sardellitti~\textit{et~al.}~\cite{mao2016dynamic} tackle computation offloading for  edge  computing in MIMO multicell system so that the optimization variables to minimize mobile devices' energy consumption include the transmit precoding matrices of mobile devices in addition to computational resources assigned by the edge cloud to the devices. Chen and Hao~\cite{chen2018task} analyze computation offloading in the framework of software defined networking (SDN), where the SDN controller decides for each offloaded task which edge cloud and how much computing resource to use.

\textbf{Intelligent reflecting surface-aided communications}. Since IRSs can be controlled to reflect incident wireless signals in a desired way, IRS-aided communications have recently received much attention in the literature~\cite{Junginproceedings,jung2018performance,huang2019Reconfigurable,fu2019intelligent,yu2019miso,nadeem2019intelligent,mishra2019channel}. The studies include analyses of data rates~\cite{Junginproceedings,jung2018performance}, optimizations of power or spectral efficiency~\cite{huang2019Reconfigurable,yu2019miso}, and channel estimation~\cite{nadeem2019intelligent,mishra2019channel}. In these studies, IRSs are also referred to as \textit{large intelligent surface}~\cite{hu2018beyond,jung2018performance}, \textit{intelligent reflecting surface}~\cite{nadeem2019intelligent,jiang2019over,qingqing2019towards}, \textit{software-defined surface}~\cite{basar2019large}, and \textit{passive intelligent mirrors}~\cite{huang2018achievable}~\cite{di2019reflection}.

We now first discuss IRS studies analyzing uplink communications, where IRSs help transmissions from mobile devices to base stations, since our work also focuses on uplinks. Jung\textit{~et~al.}~\cite{Junginproceedings,jung2018performance} analyze the impact of channel estimation errors on the uplink data rates. 

For IRSs assisting communications between mobile devices and base stations, in addition to uplink studies discussed above, downlinks are investigated in~\cite{nadeem2019large,wu2018intelligent,wu2018intelligentfull,huang2019Reconfigurable,yu2019miso}. Regarding that there seems to be more downlink studies than uplink ones, a possible explanation is that in the former case, the base station can optimize the transmit beamforming in a centralize manner.

In addition to the above settings where IRSs aid communications between mobile devices and base stations, direct communications between mobile devices and IRSs are analyzed in~\cite{jung2019performance,hu2017potential,hu2018beyond,hu2018capacity}.





%

\section{System Model} \label{sec-System}

We consider a wireless network consisting of a multi-antenna base station (BS)/access point (AP) and $K$ single-antenna mobile devices (MDs). An illustration of the system model has been given in Fig.~\ref{fig-system}. The BS/AP hosts an edge server for providing edge computing via wireless communications. The mobile devices are numbered from $1$ to $K$. For $k \in \mathcal{K}\triangleq\{1,2,\ldots, K\}$, the $k$-th mobile device has a computation task $\mathcal{J}_k : = (b_k, d_k)$ which it intends to offload to the edge server, where $b_k$ denotes the data size of the task, and $d_k$ denotes the number of CPU cycles to execute the task.

We now introduce the computing model and communication model used in this paper. Mathematical details for them are deferred to Sections~\ref{subsec-Choices-Mobile-Users} and~\ref{subsec-Transmission}.


\textbf{Computing model.}
Our computing model mainly follows that of the seminal work~\cite{chen2014decentralized,chen2015efficient} discussed in  Section~\ref{sec-Related-Work}. The same as~\cite{chen2014decentralized,chen2015efficient}, mobile devices choose computation offloading to the edge cloud or local computing on the devices by comparing the utilities under the two settings, where the utility in each setting depends on the time and energy required to complete the computation task.
 Compared with~\cite{chen2014decentralized,chen2015efficient}, our model also requires the mobile device to issue a payment to the edge cloud if a weighted function of the time and energy required to complete the computation task has  a larger value under edge computing compared with local computing.
Formal details of the computing model will be presented in Section~\ref{subsec-Choices-Mobile-Users}.

 \textbf{Communication model.} The communications between mobile devices and the BS/AP are aided by IRSs. The signals also interfere each other when arriving at the base station. Formal details of the communication model will be presented in Section~\ref{subsec-Transmission}.






\section{Optimization Problem Formulation} \label{sec-Problem-Formulation}

\subsection{A Most Generic Model: Local Computing versus Edge Computing} \label{subsec-Choices-Mobile-Users}

\textbf{Utility of a mobile device under local computing.} Below we compute the utility of the $k$-th mobile device when it completes the task $\mathcal{J}_k$ under local computing, where $k \in \mathcal{K}$.

Let the $k$-th mobile device's computation capability be $c_k^{(m)}$ CPU cycles per unit time, where the superscript ``$(m)$'' throughout the paper is used for quantities with a mobile device. Then under local computing, the
computation execution time of the task $\mathcal{J}_k$ with $d_k$ CPU cycles is
\begin{align}
t_k^{(m)} : = \frac{d_k}{c_k^{(m)}}. \label{eq-tkm}
\end{align}

Let $\mu_k$ be the energy per CPU cycle. The energy to complete the task $\mathcal{J}_k$ with $d_k$ CPU cycles is
\begin{align}
e_k^{(m)} := \mu_k d_k. \label{eq-ekm}
\end{align}

We consider that the $k$-th mobile device combines the time and energy to complete the task $\mathcal{J}_k$ in a weighted manner to quantify the cost. Specifically, with the time and energy weighted by coefficients $w_k^{(t)}$ and $w_k^{(e)}$ respectively, the combined cost for the $k$-th mobile device is $w_k^{(t)} t_k^{(m)} + w_k^{(e)} e_k^{(m)}$.
Let the benefit (without subtracting the cost) of completing the task $\mathcal{J}_k$ be $f_k(b_k, d_k)$. Then the gross utility  for the $k$-th mobile device to complete the task $\mathcal{J}_k$ under local computing is
\begin{align}
 U_k^{(m)}(b_k, d_k)& := f_k(b_k, d_k) - w_k^{(t)} t_k^{(m)} - w_k^{(e)} e_k^{(m)}.  \label{eq-lc-utilty-1}
\end{align}
Substituting Eq.~(\ref{eq-tkm}) and Eq.~(\ref{eq-ekm}) into  Eq.~(\ref{eq-lc-utilty-1}), we obtain
\begin{align}
 U_k^{(m)}(b_k, d_k)& = f_k(b_k, d_k) - w_k^{(t)} \frac{d_k}{c_k^{(m)}} - w_k^{(e)} \mu_k d_k. \label{eq-lc-utilty-2}
\end{align}

\textbf{Utility of a mobile device under edge computing.} We now compute the utility of the $k$-th mobile device when its task $\mathcal{J}_k$ is completed under edge computing, where $k \in \mathcal{K}$.

First, we analyze the time needed for the $k$-th mobile device to send the task $\mathcal{J}_k$ with $b_k$ bytes to the edge server. This depends on the transmission rate, which further depends on which of the other $K-1$ users choose edge computing and hence are transmitting. Thus, we introduce an indicator variable $a_k$ to represent the $k$-th mobile device's choice of edge computing or local computing; specifically,
\begin{subnumcases}{\hspace{-22pt}a_k \hspace{-1.5pt}=}
\hspace{-5pt}1, &\hspace{-16pt}if the $k$-th mobile device selects edge computing, \\
\hspace{-5pt}0, &\hspace{-16pt}if the $k$-th mobile device selects local computing.
\end{subnumcases}

Let $[-k]$ be the indices of mobile devices other than the $k$-th one; i.e., $[-k]\triangleq\mathcal{K}\setminus\{k\}$.
Let a vector $\boldsymbol{a}_{-k}$ represent the choices made by mobile devices in $[-k]$; i.e.,
$\boldsymbol{a}_{-k}\triangleq[a_i: i\in \mathcal{K}\setminus\{k\}]^T$.

For an IRS with $N$ IRS units, the IRS phase shift matrix as a diagonal matrix $\boldsymbol{\Phi}\in \mathbb{C}^{N \times N}$ is given by
\begin{align}
 & \boldsymbol{\Phi} := \sf{Diag}(\boldsymbol{\phi}) , \textup{ for }
\boldsymbol{\phi}\triangleq[\kappa_1e^{j \theta_1}, \ldots, \kappa_Ne^{j \theta_N}]^T ,  \label{eq-def-Phi-matrix}
\end{align}
with $\kappa_n$ and $\theta_n$ denoting the reflecting amplitude and angle of the $n$-th IRS unit respectively, for $n \in \mathcal{N}\triangleq\{1,2,\ldots, N\}$. To fully exploit the reflecting surface's ability to adjust the wireless environment, here we assume that the reflecting amplitude and reflecting angle could take values within  $[0,1]$ and $[0,2\pi)$ respectively. The data transmission rate of the $k$-th mobile device depends on $\boldsymbol{a}_{-k}$ and the IRS reflecting matrix $\boldsymbol{\Phi}$ (or the vector $\boldsymbol{\phi}$), and we denote it by  $R_k(\boldsymbol{a}_{-k}, \boldsymbol{\phi})$. The specific expression of $R_k(\boldsymbol{a}_{-k}, \boldsymbol{\phi})$ will be clear in Section~\ref{subsec-Transmission}. Then the time to transmit the data of $b_k$ bytes to the edge server is
\begin{align}
 & t_{k,send}^{(e)}(\boldsymbol{a}_{-k}, \boldsymbol{\phi}) = \frac{b_k}{R_k(\boldsymbol{a}_{-k}, \boldsymbol{\phi})}, \label{eq-time-send-cloud}
\end{align}
where the superscript ``$(e)$'' throughout the paper is used for quantities associated with the edge server.

In the transmission of the $k$-th mobile device for choosing edge computing,
let $\nu_k$ be the energy consumed per unit time. Then after time $t_{k,send}^{(e)}(\boldsymbol{a}_{-k}, \boldsymbol{\phi}) $, an amount of $\nu_k t_{k,send}^{(e)}(\boldsymbol{a}_{-k}, \boldsymbol{\phi}) $ energy is consumed. The same as~\cite{chen2015efficient}, we also consider that the $k$-th mobile device commits a tail energy of $L_k$ after transmitting the data of $b_k$ bytes. Then the total energy spent by the $k$-th mobile device is
\begin{align}
 & e_k^{(e)}(\boldsymbol{a}_{-k}, \boldsymbol{\phi}) = \nu_k t_{k,send}^{(e)}(\boldsymbol{a}_{-k}, \boldsymbol{\phi}) + L_k.
 \label{eq-energy-send-cloud}
\end{align}

After the data of $b_k$ bytes is fully transmitted to the edge server, with the edge server computation capability denoted by $c^{(e)}$ CPU cycles per unit time, the time for the edge server to complete the task $\mathcal{J}_k$ with $d_k$ CPU cycles is
\begin{align}
 & t_{k,exe}^{(e)}  = \frac{d_k}{c^{(e)}} . \label{eq-time-exe-cloud}
\end{align}

As in~\cite{chen2014decentralized,chen2015efficient}, we ignore the time required for the base station to send back the result to a mobile device. Hence, our study includes only uplinks but not downlinks in Fig.~\ref{fig-system}. Then under edge computing, the total time to complete the task $\mathcal{J}_k$ is $t_{k,send}^{(e)}(\boldsymbol{a}_{-k}, \boldsymbol{\phi}) + t_{k,exe}^{(e)} $. With $w_k^{(t)}$ and $w_k^{(e)}$ being the weights for the time and energy in the cost computation, the combined cost is $ w_k^{(t)} \cdot\left[ t_{k,send}^{(e)} (\boldsymbol{a}_{-k}, \boldsymbol{\phi}) + t_{k,exe}^{(e)}  \right] + w_k^{(e)} \cdot e_k^{(e)}(\boldsymbol{a}_{-k}, \boldsymbol{\phi})$. Under edge computing, we consider that the $k$-th mobile device also needs to have a payment of $P_k$ to the edge server. Then with the benefit (without subtracting the cost) of completing the task $\mathcal{J}_k$ being $f_k(b_k, d_k)$, the gross utility  for the $k$-th mobile device to complete the task $\mathcal{J}_k$ under edge computing is
\begin{align}
 & U_k^{(e)} (\boldsymbol{a}_{-k}, \boldsymbol{\phi}, b_k, d_k) \nonumber \\ &= f_k(b_k, d_k) -  w_k^{(t)} \cdot\left[ t_{k,send}^{(e)} (\boldsymbol{a}_{-k}, \boldsymbol{\phi}) + t_{k,exe}^{(e)}  \right] \nonumber \\ & \quad - w_k^{(e)} \cdot e_k^{(e)}(\boldsymbol{a}_{-k}, \boldsymbol{\phi}) - P_k . \label{eq-utility-cloud-1}
\end{align}

Substituting Eq.~(\ref{eq-time-send-cloud})~(\ref{eq-energy-send-cloud})~(\ref{eq-time-exe-cloud}) into Eq.~(\ref{eq-utility-cloud-1}), we obtain
\begin{align}
 & U_k^{(e)} (\boldsymbol{a}_{-k}, \boldsymbol{\phi}, b_k, d_k) \nonumber \\ &= f_k(b_k, d_k) -   w_k^{(t)}  \cdot \left[\frac{b_k}{R_k(\boldsymbol{a}_{-k}, \boldsymbol{\phi})} + \frac{d_k}{c^{(e)}} \right] \nonumber \\ & \quad -  w_k^{(e)} \cdot \left[\frac{\nu_k b_k}{R_k(\boldsymbol{a}_{-k}, \boldsymbol{\phi})} + L_k\right] - P_k . \label{eq-utility-cloud-2}
\end{align}

In our studied model, for the $k$-th mobile device, if the gross utility obtained under edge computing is at least the gross utility obtained under local computing, then the $k$-th mobile device will have $a_k =1$ to choose edge computing over local computing; otherwise, $a_k =0$. Formally,
\begin{align}
a_k = \begin{cases}
1, & \textup{if $U_k^{(e)} (\boldsymbol{a}_{-k}, \boldsymbol{\phi}, b_k, d_k) \geq  U_k^{(m)}(b_k, d_k)$,} \\
0, & \textup{otherwise.}
\end{cases} \label{eq-ak10v1}
\end{align}

From Eq.~(\ref{eq-lc-utilty-2}) and Eq.~(\ref{eq-utility-cloud-2}),
the inequality $U_k^{(e)} (\boldsymbol{a}_{-k}, \boldsymbol{\phi}, b_k, d_k) \geq U_k^{(m)}(b_k, d_k)$ is equivalent to
\begin{align}
 P_k \leq \, & f_k(b_k, d_k) - w_k^{(t)}  \cdot \left[\frac{b_k}{R_k(\boldsymbol{a}_{-k}, \boldsymbol{\phi})} + \frac{d_k}{c^{(e)}} \right] \nonumber  \\ &  - w_k^{(e)} \cdot \left[\frac{\nu_k b_k}{R_k(\boldsymbol{a}_{-k}, \boldsymbol{\phi})} + L_k\right]  \nonumber  \\ &  - \left[f_k(b_k, d_k) - w_k^{(t)} \frac{d_k}{c_k^{(m)}} - w_k^{(e)} \mu_k d_k\right]. \label{eq-Pk-inequ}
\end{align}

Defining $C_k$ and $A_k(b_k)$ by
\begin{align}
C_k  :=  w_k^{(t)} \frac{d_k}{c_k^{(m)}} + w_k^{(e)} \mu_k d_k - \frac{w_k^{(t)} d_k}{c^{(e)}} - w_k^{(e)} L_k  . \label{eq-def-Ck}
\end{align}
and
\begin{align}
A_k(b_k) := w_k^{(t)} b_k+w_k^{(e)}\nu_k b_k, \label{eq-def-Ak}
\end{align}
we write the above Inequality~(\ref{eq-Pk-inequ}) as
\begin{align}
 P_k \leq C_k - \frac{A_k(b_k)}{R_k(\boldsymbol{a}_{-k}, \boldsymbol{\phi})},\label{eq-Pk-inequ2}
\end{align}
which converts Eq.~(\ref{eq-ak10v1}) into
\begin{align}
a_k = \begin{cases}
1, & \textup{if $P_k \leq C_k - \frac{A_k(b_k)}{R_k(\boldsymbol{a}_{-k}, \boldsymbol{\phi})}$,} \\
0, & \textup{otherwise.}
\end{cases} \label{eq-ak10v2}
\end{align}

From Eq.~(\ref{eq-ak10v2}), we have the following:
\begin{itemize}[leftmargin=10pt]
\item If $C_k - \frac{A_k(b_k)}{R_k(\boldsymbol{a}_{-k}, \boldsymbol{\phi})} \geq 0$, the edge server can set $P_k \in [0, C_k - \frac{A_k(b_k)}{R_k(\boldsymbol{a}_{-k}, \boldsymbol{\phi})}]$, and the $k$-th mobile device will have $a_k =1$ to choose edge computing over local computing.
\item If $C_k - \frac{A_k(b_k)}{R_k(\boldsymbol{a}_{-k}, \boldsymbol{\phi})} < 0$, in case that the edge server does not accept a negative payment from the $k$-th mobile device (i.e., the edge server does not compensate the mobile device for choosing edge computing), then the $k$-th mobile device will have $a_k =0$ to choose local  computing over edge computing, and the edge server will receive no payment from the $k$-th mobile device so that $P_k =0$.
\end{itemize}

Besides, the information rate for each MD should be no smaller than a requirement $r_k$ to ensure that each MD could fulfill a basic communication and/or computation task.  

From the above analysis, the maximal payment that the edge server can get from the $k$-th mobile device is \mbox{$\max\left\{C_k - \frac{A_k(b_k)}{R_k(\boldsymbol{a}_{-k}, \boldsymbol{\phi})},0\right\}$.}  Then to maximize the total payment from all $K$ mobile devices, the edge server solves the following optimization Problem~(P1):
\begin{subequations} \label{opt-p1}
\begin{alignat}{2}
\textup{(P1):}\max_{\boldsymbol{\phi}}&\  \sum_{k=1}^K \max\left\{C_k - \frac{A_k(b_k)}{R_k(\boldsymbol{a}_{-k}, \boldsymbol{\phi})},0\right\} \label{opt-p1-objective} \\
\sf{s.t.} &\ a_k = \begin{cases}
1, &  \textup{if $C_k - \frac{A_k(b_k)}{R_k(\boldsymbol{a}_{-k}, \boldsymbol{\phi})} \geq 0$}, \nonumber \\
0, & \textup{otherwise},
\end{cases} \forall k \in\mathcal{K}, \label{opt-p1-constraint1} \\ 
& R_k(\boldsymbol{a}_{-k}, \boldsymbol{\phi})\geq r_k, \forall k, \\
& | \boldsymbol{\phi}_n |\leq1, \forall n\in\mathcal{N}.\label{opt-p1-constraint2}
\end{alignat}
\end{subequations}
The above constraint~(\ref{opt-p1-constraint2}) holds since $\boldsymbol{\phi}_n=\kappa_ne^{j \theta_n}$ from Eq.~(\ref{eq-def-Phi-matrix}).

\subsection{A Simplified Case} \label{subsec-BS-profit}

Maximizing the non-concave objective function in~(\ref{opt-p1-objective}) even without considering the constraints~(\ref{opt-p1-constraint1}) and (\ref{opt-p1-constraint2}) is difficult. Hence, we consider that if $C_k - \frac{A_k(b_k)}{R_k(\boldsymbol{a}_{-k}, \boldsymbol{\phi})}<0$, the edge server accepts a negative payment from the $k$-th mobile device (i.e., the edge server compensates the mobile device for choosing edge computing). Then regardless of the relationship between $C_k - \frac{A_k(b_k)}{R_k(\boldsymbol{a}_{-k}, \boldsymbol{\phi})} $ and $0$, the maximal payment that the edge server can get from the $k$-th mobile device is always $C_k - \frac{A_k(b_k)}{R_k(\boldsymbol{a}_{-k}, \boldsymbol{\phi})}$. Also, the $k$-th mobile device for each $k\in\mathcal{K}$ will always have $a_k =1$ to choose edge computing over local computing. Now with $a_i=1$ for $i\in\mathcal{K}$, $\boldsymbol{a}_{-k}$ becomes a vector of all $1$s, and we write $R_k(\boldsymbol{a}_{-k}, \boldsymbol{\phi})$ as $R_k(\boldsymbol{\phi})$ for notation simplicity. Thus, to maximize the total payment from all $K$ mobile devices, the edge server solves the following optimization Problem~(P2):
\begin{subequations}\label{opt-p2}
\begin{alignat}{2}
\textup{(P2):} \quad \max_{\boldsymbol{\phi}} & \qquad \sum_{k=1}^K \left[C_k - \frac{A_k(b_k)}{R_k(  \boldsymbol{\phi})}\right] \label{opt-p2-objective} \\
\textup{s.t.} & \ R_k(\boldsymbol{\phi})\geq r_k, \forall k\in\mathcal{K}, \label{opt-p2-rate_cnstr} \\
&\ \ |\boldsymbol{\phi}_n|\leq1, \forall n\in\mathcal{N}.\label{opt-p2-constraint2}
\end{alignat}
\end{subequations}
Later we will show that the problem (P2) can be efficiently solved. Hence, Problem~(P2) of (\ref{opt-p2}) significantly reduces the difficulty of solving Problem~(P1) of (\ref{opt-p1}).

Problem~(P2) provides a lower bound for Problem~(P1). Moreover, considering Problem~(P2) also has practical interest. As discussed above, we   consider that  the edge server is willing to compensate a mobile device for choosing edge computing if providing edge computing to the mobile device for  some computation task does not bring immediate earning. Doing so may encourage the mobile device to adopt edge computing for future computation tasks, and the edge server can profit from the mobile device in these future tasks. In addition, the edge server's compensation may help the edge server to increase the user base of its edge computing and hence improve the long-term earning.


\subsection{Signal Model}
\label{subsec-Transmission}

In this paper we consider deploying the IRS to assist a typical wireless communication system, where a multi-antenna AP is communicating with $K$ single-antenna MDs. Suppose that the AP has $M$ antennas and the IRS equipment has $N$ reflecting elements. We use $\boldsymbol{h}_{\textup{r},k}^H \in \mathbb{C}^{1 \times N}$ to denote the wireless channel from the $k$-th MD to the IRS. Let $\boldsymbol{G} \in \mathbb{C}^{M \times N} $ be the channel from the IRS to the AP. Define $\boldsymbol{h}_{\textup{d},k}^H\in \mathbb{C}^{1 \times M} $ to be the channel from the $k$-th MD to the AP. 

Here we consider the system experiences quasi-static flat-fading channel. Therefore the AP can obtain the channel status information (CSI) $\boldsymbol{h}_{\textup{r},k}^H, \boldsymbol{G} ,\boldsymbol{h}_{\textup{d},k}^H $ via the standard channel estimation and feedback technique (e.g. in a classical time-division duplexing (TDD) system).  
Based on knowledge of $\boldsymbol{h}_{\textup{r},k}^H, \boldsymbol{G} ,\boldsymbol{h}_{\textup{d},k}^H$, the AP manipulates the amplitude and phase of the IRS elements to improve its earnings. That is AP tries to solve the Problem~(P2) with respect to $\boldsymbol{\phi}$. 

In view of $\boldsymbol{\Phi}=\textup{diag}(\boldsymbol{\phi})$, we define the effective channel $\boldsymbol{h}_k(\boldsymbol{\phi})\in \mathbb{C}^{M \times 1}$ of the $k$-th MD by
\begin{align}
 &\boldsymbol{h}_k(\boldsymbol{\phi}):=  \boldsymbol{G} \textup{diag}(\boldsymbol{\phi}) \boldsymbol{h}_{\textup{r},k} + \boldsymbol{h}_{\textup{d},k}. \label{eq-def-hk}
\end{align}
The received signal at AP can be represented as 
\begin{align}
\boldsymbol{r}(\boldsymbol{\phi}) = \sum_{j}\sqrt{q}_j\boldsymbol{h}_j(\boldsymbol{\phi})s_j+\boldsymbol{n},  \label{eq-rcvd_sig}
\end{align}
where the vector $\boldsymbol{n}\in\mathbb{C}^M$ denotes the local thermal noise at the receiver and is usually modeled as Gaussian distribution $\mathcal{CN}(\boldsymbol{0},\sigma^2\boldsymbol{I}_M)$.   


Assume that linear receiver $\boldsymbol{u}_k^H\in \mathbb{C}^{1\times M}$ is utilized at the AP to decode signals associated with the $k$-th MD. Then the signal-to-interference-plus-noise ratio (SINR) is a function of $\boldsymbol{\phi}$ given as
\begin{align}
\gamma_k\big(\boldsymbol{\phi}\big) = \frac{q_k|\boldsymbol{u}_k^H \boldsymbol{h}_k(\boldsymbol{\phi})|^2}{\boldsymbol{u}_k^H \boldsymbol{W}_k( \boldsymbol{\phi}) \boldsymbol{u}_k }. \label{eq-gamma-k}
\end{align}
with interference-plus-noise spatial covariance matrix
\begin{align}
\boldsymbol{W}_k( \boldsymbol{\phi}) = \sigma^2 \boldsymbol{I}_{M}+\sum_{i\neq k} q_i \boldsymbol{h}_i(\boldsymbol{\phi}) \boldsymbol{h}_i^H(\boldsymbol{\phi}). \label{eq-Zk-matrix}
\end{align}
To improve receiving quality, the optimal $\boldsymbol{u}_k^\star$ \cite{schubert2005iterative,monzingo1980introduction} 
\begin{align}
\boldsymbol{u}_k^\star = \frac{[{\boldsymbol{W}_k( \boldsymbol{\phi})}]^{-1}  \boldsymbol{h}_k(\boldsymbol{\phi})}{\|[{\boldsymbol{W}_k( \boldsymbol{\phi})}]^{-1} \boldsymbol{h}_k(\boldsymbol{\phi})\|_2}  , \label{eq-mu-k-optimal}
\end{align}
should be  used to yield the maximum $\gamma_k^{\star}\big(\boldsymbol{\phi}\big)$ given as 
\begin{align}
\gamma_k^\star\big(\boldsymbol{\phi}\big)   = q_k\boldsymbol{h}_k^H(\boldsymbol{\phi}) [{\boldsymbol{W}_k( \boldsymbol{\phi})}]^{-1}  \boldsymbol{h}_k(\boldsymbol{\phi}). \label{eq-gamma-k-optimal}
\end{align}

Assuming that unit bandwidth is used, then by Shannon's formula~\cite{shannon1948mathematical} the allowed information rate of the $k$-th MD $R_k( \boldsymbol{\phi})$ is evaluated as \footnote{Here we somewhat abuse the notation $R_k(\cdot)$. Our $R_k\big(\boldsymbol{\phi}\big)$ is actually defined as the information rate associated with SINR maximizing receiving vector, not for generic receiving vectors.}  
\begin{align}
 &R_k( \boldsymbol{\phi}) = \log(1+\gamma_k^{\star}(\boldsymbol{\phi}) )  . \label{eq-Rk-rate}
\end{align}
Considering result in (\ref{eq-gamma-k-optimal}) we obtain  
\begin{align}
R_k( \boldsymbol{\phi})& =\log(1+ q_k\boldsymbol{h}_k^H(\boldsymbol{\phi}) [{\boldsymbol{W}_k( \boldsymbol{\phi})}]^{-1}  \boldsymbol{h}_k(\boldsymbol{\phi})) \label{eq-obj-1-Rk} .
\end{align}

\subsection{Problem Formulation} \label{subsec-optimizing-Mobile-IRS}

Putting the above discussions together, our goal is to solve (P2) in (\ref{opt-p2}), where $R_k( \boldsymbol{\phi})$ is given by (\ref{eq-obj-1-Rk}).

In~(\ref{opt-p2-objective}), since $b_k$ does not depend on the optimization variable $\boldsymbol{\phi}$, we just abbreviate $A_k(b_k)$ to $A_k$ in the following discussion. Maximizing the term in~(\ref{opt-p2-objective}) is the same as minimizing $\sum_{k=1}^K  \frac{A_k}{R_k(\boldsymbol{\phi}) }$, so we can write Problem~(P2) as the following Problem~(P3):
\begin{subequations}
\begin{alignat}{2}
\textup{(P3):} \quad \min_{\boldsymbol{\phi}} & \qquad \sum_{k=1}^K  \frac{A_k}{R_k(\boldsymbol{\phi}) } \\
\textup{s.t.} & \ R_k(\boldsymbol{\phi})\geq r_k, \forall k\in\mathcal{K}, \\
&\ \ |\boldsymbol{\phi}_n|\leq1, \forall k\in\mathcal{N}.
\end{alignat}
\end{subequations}
We will discuss the feasibility and solution of (P3) in the subsequent sections in details.

\section{Feasibility Problem} \label{Feasibility-Check}

The problem (P3) is difficult due to both of its complex objective as well as the group of rate constraints, which are all nonconvex. Putting aside the highly nonconvex objective, one first question comes to our mind is the feasibility of the problem. That is, with an arbitrary set of rate constraints $\{r_k\}$ assigned, is the problem (P3) feasible? Determining the feasibility of (P3) is itself hard but at the same time highly meaningful. Since it is the base for scheming a reasonable rate requirement and serves as a starting point to solve (P3) (as will become clear in the next section). In this section we develop an iterative evaluation procedure, which i)
can serve as a sufficient condition for the feasibility of (P3) and ii) can provide a feasible solution when (P3) is identified as feasible. 

Noticing the relation in (\ref{eq-Rk-rate}), the feasibility check of (P3), can be equivalently written as the following problem: 
\begin{subequations}
\begin{alignat}{2}
\textup{(P4):} \quad \sf{Find}\ & {\boldsymbol{\phi}} \\
\sf{s.t.} & \gamma_k^{\star}(\boldsymbol{\phi})\geq e^{r_k}-1, \forall k\in\{1,\cdots,K\}, \\
&\ \ |\boldsymbol{\phi}_n|\leq1, \forall n\in\{1,\cdots,N\}.
\end{alignat}
\end{subequations}

To further simplify the above SINR constraints, which have fractional nature, we consider a closely related metric---mean square error (MSE), which is defined as 
\begin{align}
\varepsilon_k(\boldsymbol{\phi},\boldsymbol{w}_k) \triangleq \mathbb{E}\Big\{\big|s_k\!-\!\boldsymbol{w}_k^H\big(\sum_{j}\sqrt{q}_j\boldsymbol{h}_js_j\!+\!\boldsymbol{n}\big)\big|^2\Big\}\label{MSE_def}
\end{align}
where the vector $\boldsymbol{w}_k$ is the linear receiver utilized at AP to suppress the interference and noise. The expectation in (\ref{MSE_def}) is taken over the distribution of noise $\boldsymbol{n}$ and information symbols $\{s_j\}$. When linear receiver is deployed, the following identity connects the minimal MSE $\varepsilon^{\star}_k(\boldsymbol{\phi})$ and maximal SINR $\gamma^{\star}_k(\boldsymbol{\phi})$:
\begin{align}
\varepsilon^{\star}_k(\boldsymbol{\phi}) = \frac{1}{1+\gamma^{\star}_k(\boldsymbol{\phi})}. \label{mse_sinr}
\end{align}
In fact, it can be proved that optimal linear MSE receiver is given as 
\begin{align}
\boldsymbol{w}_k^{\star} = \widetilde{\boldsymbol{W}}(\boldsymbol{\phi})^{-\!1}\sqrt{q_k}\boldsymbol{h}_k(\boldsymbol{\phi}), \label{eq:w_mmse_lin}
\end{align}
with $\widetilde{\boldsymbol{W}}(\boldsymbol{\phi})$ being
\begin{align}
\widetilde{\boldsymbol{W}}(\boldsymbol{\phi}) = \sigma^2\boldsymbol{I}_M+\sum_{j}q_{i}\boldsymbol{h}_j(\boldsymbol{\phi})\boldsymbol{h}_j(\boldsymbol{\phi})^H.\label{W_tilde}
\end{align}
and the minimal $\varepsilon^{\star}_k(\boldsymbol{\phi})$ can be readily obtained as $\varepsilon^{\star}_k(\boldsymbol{\phi})=1\!-\!q_k\boldsymbol{h}_k(\boldsymbol{\phi})^H\widetilde{\boldsymbol{W}}_k(\boldsymbol{\phi})^{-\!1}\boldsymbol{h}_k(\boldsymbol{\phi})$. Combined with (\ref{eq-gamma-k-optimal}), (\ref{mse_sinr}) can be verified. 

Based on the above observation in (\ref{mse_sinr}), (P4) can equivalently transformed into 
\begin{subequations}
\begin{alignat}{2}
\textup{(P5):} \quad \sf{Find}\ & {\boldsymbol{\phi}} \\
\sf{s.t.} &\ \varepsilon_k^{\star}(\boldsymbol{\phi})\leq \frac{1}{1+e^{r_k}-1}=e^{-r_k}, \forall k\in\mathcal{K}, \\
&\ \ |\boldsymbol{\phi}_n|\leq1, \forall n\in\mathcal{N}.
\end{alignat}
\end{subequations}

The above optimization problem can be further transformed as follows
\begin{subequations}
\begin{alignat}{2}
\textup{(P6):} \quad \sf{Find}\ & \boldsymbol{\phi},\{\boldsymbol{w}_k\}  \\
\sf{s.t.} &\ \varepsilon_k(\boldsymbol{\phi},\{\boldsymbol{w}_k\})\leq e^{-r_k}, \forall k\in\mathcal{K}, \\
&\ \ |\boldsymbol{\phi}_n|\leq1, \forall n\in\mathcal{N}.
\end{alignat}
\end{subequations}
The equivalence between (P5) and (P6) can be seen by noticing the fact that $\varepsilon_k^{\star}(\boldsymbol{\phi})=\min_{\boldsymbol{w}_k}\varepsilon_k(\boldsymbol{\phi},\boldsymbol{w}_k)$ and the optima is achieved when choosing $\boldsymbol{w}_k=\boldsymbol{w}_k^{\star}$ given in (\ref{eq:w_mmse_lin}). In fact, if $\varepsilon_k^{\star}(\boldsymbol{\phi})\leq e^{r_k}$, then $\varepsilon_k(\boldsymbol{\phi},\boldsymbol{w}_k^{\star})=\varepsilon_k^{\star}(\boldsymbol{\phi})\leq e^{r_k}$. Conversely, if there exists $(\boldsymbol{\phi},\boldsymbol{w}_k)$ such that $\varepsilon_k(\boldsymbol{\phi},\boldsymbol{w}_k)\leq e^{r_k}$, then $\varepsilon_k^{\star}(\boldsymbol{\phi})\leq\varepsilon_k(\boldsymbol{\phi},\boldsymbol{w}_k)\leq e^{-r_k}$.

Now we consider the following optimization problem  
\begin{subequations}
\begin{alignat}{2}
\textup{(P7):} \quad \min_{\boldsymbol{\phi},\{\boldsymbol{w}_k\},\alpha} \ &\alpha \\
\sf{s.t.} &\ e^{r_k}\varepsilon_k(\boldsymbol{\phi},\boldsymbol{w}_k)\leq\alpha, \forall k\in\mathcal{K}, \\
&\ \ |\boldsymbol{\phi}_n|\leq1, \forall n\in\mathcal{N}.
\end{alignat}
\end{subequations}
If the optimal value of (P7) is no greater than $1$, then (P6), and consequently the original problem (P3), is feasible. Otherwise the problem is infeasible. The intention behind the transformation from (P4) to (P7) lies in the fact that $\varepsilon_k(\boldsymbol{\phi},\boldsymbol{w}_k)$ has a quadratic form and is bi-convex with respect to $\boldsymbol{\phi}$ and $\{\boldsymbol{w}_k\}$, which is much simpler to tackle compared to $\gamma_k^{\star}(\boldsymbol{\phi})$.

Since the problem is still nonconvex with respect to $\boldsymbol{\phi}$ and $\{\boldsymbol{w}_k\}$ jointly, we can use block coordinate descent (BCD) method to solve it. Specifically we can alternatively update $(\boldsymbol{\phi},\alpha)$ and $(\{\boldsymbol{w}_k\},\alpha)$. When $\boldsymbol{\phi}$ is fixed, the optimal $\{\boldsymbol{w}_k\}$ is given in (\ref{eq:w_mmse_lin}) as discussed previously and $\alpha$ can be accordingly determined. When $\{\boldsymbol{w}_k\}$ are fixed, $\boldsymbol{\phi}$ should be obtained via solving the following problem:
\begin{subequations}
\begin{alignat}{2}
\!\!\!\!\!\!\textup{(P8):} \min_{\boldsymbol{\phi},\alpha} \ &\alpha \\
\sf{s.t.} &\ \boldsymbol{\phi}^H\boldsymbol{Q}_k\boldsymbol{\phi}+2\mathrm{Re}\{\boldsymbol{q}_k^H\boldsymbol{\phi}\}+d_k\leq\alpha, \forall k\in\mathcal{K}, \\
&\ \ |\boldsymbol{\phi}_n|\leq1, \forall n\in\mathcal{N}.
\end{alignat}
\end{subequations}
with the parameters defined as follows:
\begin{subequations}
\begin{align}
\!\!\!\!\boldsymbol{F}_k\triangleq &\ \ \boldsymbol{G}\sf{Diag}(\boldsymbol{h}_{\rm{r},k}), \\ 
\!\!\!\!\boldsymbol{Q}_k\triangleq & e^{r_k}\sum_{j}q_j\boldsymbol{F}_j^H\boldsymbol{w}_k\boldsymbol{w}_k^H\boldsymbol{F}_j, \label{eq_cnstr_Q}\\ 
\boldsymbol{q}_k\triangleq & e^{r_k}\sum_{j}q_j\boldsymbol{w}_k^H\boldsymbol{h}_{\rm{d},j}\boldsymbol{F}_j^H\boldsymbol{w}_k\!-\!e^{r_k}\sqrt{q_k}\boldsymbol{F}_k^H\boldsymbol{w}_k \\
d_k \triangleq & e^{r_k}\big(\sum_{j}q_j\big|\boldsymbol{w}_k^H\boldsymbol{h}_{\rm{d},j}\big|^2\!\!-\!\!2\sqrt{q_k}\mathrm{Re}\{\boldsymbol{w}_k^H\boldsymbol{h}_{\rm{d},k}\} \\
& \quad\quad+\!\sigma^2\|\boldsymbol{w}_k\|_2^2\!+\!1\big).\nonumber
\end{align}
\end{subequations}

According to (\ref{eq_cnstr_Q}), $\boldsymbol{Q}_k$ is obvioulsy positive-semidefinite $\forall k$ and therefore (P8) is convex. In fact (P8) can be transformed into standard second-order-cone-programming (SOCP) problem and can be efficiently solved via standard numerical solver like CVX.

Since the BCD procedure alternatively updates $\boldsymbol{\phi}$ and $\{\boldsymbol{w}_k\}$ minimizing $\alpha$, $\alpha$ monotonically decreases until convergence. Once the value of $\alpha$ is found to be smaller than $1$, then the original problem is feasible and a feasible $\boldsymbol{\phi}$ has been found and BCD procedure could be stop. Otherwise, $\alpha$ converges to a value larger than $1$, and we have to claim that the problem is infeasible. The above discussion is summarized in Alg.~\ref{Alg-feas-prob}. 

It should be noted that (P7) is itself nonconvex and its global optimal solution can hardly be obtained. Alg.~\ref{Alg-feas-prob} finds a suboptimal solution of (P7). Therefore Alg.~\ref{Alg-feas-prob} finding an objective value no greater than $1$ is a sufficient condition for feasibility of the original problem.

\begin{algorithm}[!t]
\caption{Feasibility Check Algorithm} \label{Alg-feas-prob}
\begin{algorithmic}[1]
\STATE Randomly initialize $\boldsymbol{\phi}^{(0)}$ with $\boldsymbol{\phi}^{(0)}_n=\kappa_n e^{j\theta_n}$ with $\kappa_n^2$ and $\theta_n$ uniformly distributed among $[0,1]$ and $[0, 2\pi)$ respectively, $\forall n\in\mathcal{N}$;
\STATE Initialize $\{\boldsymbol{w}_k^{(0)}\}$ via (\ref{eq:w_mmse_lin});
\STATE Initialize $\alpha^{(0)}:=\max_{k}\varepsilon_k(\boldsymbol{\phi}^{(0)},\boldsymbol{w}_k^{(0)})$;
\REPEAT
\STATE update $\boldsymbol{w}_k^{(t+1)}$ via (\ref{eq:w_mmse_lin});
\STATE update $\alpha^{(t+1)}$ and $\boldsymbol{\phi}^{(t+1)}$ by solving (P8);
\STATE $t++$;
\UNTIL{\emph{convergence or} $\alpha^{(t)}\leq1$}
\IF{$\alpha^{(t)}<=1$}
\STATE claim {\sf{Feasible}}, output $\boldsymbol{\phi}^{(t)}$ as feasible solution;
\ELSE
\STATE claim {\sf{Infeasible}}
\ENDIF
\end{algorithmic}
\end{algorithm}

\section{Solving the Optimization} \label{sec-Analysis-Algorithms}

In this section we develop an algorithm to solve the Problem~(P3), whose objective has a sum-of-ratios form and is difficult to solve. Based on the discussion in previous section, we assume that the problem (P3) is feasible and we can find a feasible solution of (P3). 

To attack (P3), we firstly introduce the an equivalent transformation of the rate function $R_k(\boldsymbol{\phi})$ \cite{Cioffi_wmmse, QingjiangShi_wmmse}, which will significantly simplify the optimization procedure, as will be shown later. The following identities hold:
\begin{subequations}
\label{wmmse_transform}
\begin{align}
&\!\!\!\!\!\!R_k(\boldsymbol{\phi})=\log(1+\gamma_k^\star(\boldsymbol{\phi}))\\
=&-\log\Big[\big(q_k^{-1}+\boldsymbol{h}_k^H\boldsymbol{W}_k^{-1}\boldsymbol{h}_k\big)^{-\!1}\Big]+\log(q_k))\\
\stackrel{(a)}{=}&\max_{\varpi_k\geq0}\Big\{\log(\varpi_k)-\varpi_k\Big[\big(q_k^{-1}+\boldsymbol{h}_k^H\boldsymbol{W}_k^{-1}\boldsymbol{h}_k\big)^{-1}\Big]\nonumber\\
&\quad\qquad+1+\log(q_k)\Big\}\label{wmmse_1st}\\
\stackrel{(b)}{=}&\max_{\varpi_k\geq0,\boldsymbol{v}_k}\Big\{-\varpi_k\Big[q_k\big|1-\boldsymbol{v}_k^H\boldsymbol{h}_k\big|^2+\boldsymbol{v}_k^H\boldsymbol{W}_k\boldsymbol{v}_k\Big]\nonumber\\
&\qquad\qquad+\log(\varpi_k)+1+\log(q_k)\Big\}\label{wmmse_2nd}\\
\triangleq&\max_{\varpi_k\geq0,\boldsymbol{v}_k}\widetilde{R}_k\big(\boldsymbol{\phi},\boldsymbol{v}_k,\varpi_k\big).
\end{align}
\end{subequations}
In the above step (a) can be readily verified by realizing the function in the maximization is concave with respect to $\varpi_k$ and its optima is obtained via 
\begin{align}
\varpi_k^\star= q_k^{-1}+\boldsymbol{h}_k(\boldsymbol{\phi})^H\boldsymbol{W}_k(\boldsymbol{\phi})^{-1}\boldsymbol{h}_k(\boldsymbol{\phi}).\label{w_opt}
\end{align}
The transformation (b) holds since the function in (\ref{wmmse_2nd}) is quadratic concave in $\boldsymbol{v}_k$ and its optima is achieved at
\begin{align}
\boldsymbol{v}_k^\star=\boldsymbol{W}_k(\boldsymbol{\phi})^{-1}q_k\boldsymbol{h}_k(\boldsymbol{\phi}).\label{v_opt}
\end{align}

Abbreviating the notations $\boldsymbol{\varpi}\triangleq[\varpi_1,\cdots,\varpi_K]$ and $\boldsymbol{V}\triangleq[\boldsymbol{v}_1,\cdots,\boldsymbol{v}_K]$, we transform the problem (P3) as follows
\begin{subequations}
\begin{alignat}{2}
\textup{(P9):} \quad \min_{\boldsymbol{\phi},\boldsymbol{V},\boldsymbol{\varpi}} & \qquad \sum_{k=1}^K  \frac{A_k}{\widetilde{R}_k(\boldsymbol{\phi},\boldsymbol{v}_k,\varpi_k)} \\
\textup{s.t.} & \ \widetilde{R}_k(\boldsymbol{\phi},\boldsymbol{v}_k,\varpi_k))\geq r_k, \forall k\in\mathcal{K}, \\
&\ \ |\boldsymbol{\phi}_n|\leq1, \forall n\in\mathcal{N}.
\end{alignat}
\end{subequations}
 

Introducing an intermediate variable $\boldsymbol{\mu}\triangleq[\mu_1, \ldots, \mu_K]$, (P9) can be equivalently written as
\begin{subequations}
\begin{alignat}{2}
\textup{(P10):}\qquad \min_{\boldsymbol{\phi},\boldsymbol{V},\boldsymbol{\varpi},\boldsymbol{\mu}}& \sum_{k=1}^K  \mu_k     \\
~\hspace{-37pt}\textup{s.t.}&\ \frac{A_k}{\widetilde{R}_k(\boldsymbol{\phi},\boldsymbol{v}_k,\varpi_k)} \leq \mu_k, \forall k\in\mathcal{K},  \label{p11_frac_cnstr}\\
& \ \widetilde{R}_k(\boldsymbol{\phi},\boldsymbol{v}_k,\varpi_k)\geq r_k, \forall k\in\mathcal{K}, \label{p11_rate_cnstr}\\
&\ |\boldsymbol{\phi}_n|\leq1, \forall n\in\mathcal{N}, \label{phi_cnstr}.
\end{alignat}
\end{subequations}

To solve (P10), we have the following lemmas, which are proved in the Appendix of the online full version \cite{MEC_IRS_fullpaper} (i.e., this paper):
\begin{lem} \label{lem-part1}
If $(\boldsymbol{\phi}^\star,\boldsymbol{V}^\star,\boldsymbol{\varpi}^\star,\boldsymbol{\mu}^\star)$ is a solution to (P10), then there exists $\boldsymbol{\lambda}^\star=[\lambda_1^\star, \ldots, \lambda_K^\star]$ such that $(\boldsymbol{\phi}^\star,\boldsymbol{V}^\star,\boldsymbol{\varpi}^\star)$ satisfies the Karush-Kuhn-Tucker (KKT) condition of the following problem (P11) when $\boldsymbol{\lambda}=\boldsymbol{\lambda}^\star$ and $\boldsymbol{\mu}=\boldsymbol{\mu}^\star$:
\begin{subequations}
\begin{alignat}{2}
\textup{(P11):}\min_{\boldsymbol{\phi},\boldsymbol{V},\boldsymbol{\varpi}}&\ \quad \sum_{k=1}^K\lambda_k\big(A_k\!-\!\mu_k \widetilde{R}_k(\boldsymbol{\phi},\boldsymbol{v}_k,\varpi_k)\big)\label{opt-p5-obj} \\
\rm{s.t.} & \ \widetilde{R}_k(\boldsymbol{\phi},\boldsymbol{v}_k,\varpi_k)\geq r_k, \forall k\in\mathcal{K}, \label{reduced_func_rate_cnstr}\\
& |\boldsymbol{\phi}_n|\leq1, \forall n\in\mathcal{N} \label{reduced_func_model_cnstr}.
\end{alignat}
\end{subequations}
Also, we have
\begin{align}
\lambda_k^\star \widetilde{R}_k(\boldsymbol{\phi}^\star,\boldsymbol{v}_k^\star,\varpi_k^\star)&=1,  \forall k\in\mathcal{K}, \label{lem-part1-lambda}\\
\mu_k^\star \widetilde{R}_k(\boldsymbol{\phi}^\star,\boldsymbol{v}_k^\star,\varpi_k^\star)&=A_k ,  \forall k\in\mathcal{K}. \label{lem-part1-gamma}
\end{align}
\end{lem}

\begin{lem} \label{lem-part2}
If both \ding{172} and \ding{173} below hold:
\begin{itemize}
\item[\ding{172}]  $(\boldsymbol{\phi}^\star,\boldsymbol{V}^\star,\boldsymbol{\varpi}^\star)$ solves (P11) when $\boldsymbol{\lambda}=\boldsymbol{\lambda}^\star$ and $\boldsymbol{\mu}^\star=\boldsymbol{\mu}^\star$;
\item[\ding{173}] $(\boldsymbol{\phi}^\star,\boldsymbol{V}^\star,\boldsymbol{\varpi}^\star)$, $\boldsymbol{\lambda}^\star$, and $\boldsymbol{\mu}^\star$ satisfy (\ref{lem-part1-lambda}) and (\ref{lem-part1-gamma}),
\end{itemize}
then
$(\boldsymbol{\phi}^\star,\boldsymbol{V}^\star,\boldsymbol{\varpi}^\star,\boldsymbol{\mu}^\star)$ satisfies the KKT condition of (P10).
\end{lem}

According to Lemmas~\ref{lem-part1} and~\ref{lem-part2}, we can solve (P10) via tackling (P11). Specifically, we adopt the approach of~\cite{Jonga-Sum-of-Ratios} to find $(\boldsymbol{\phi}^\star,\boldsymbol{V}^\star,\boldsymbol{\varpi}^\star)$, $\boldsymbol{\lambda}^*$, and $\boldsymbol{\mu}^*$ satisfying \ding{172} and \ding{173} of Lemma~\ref{lem-part2} via alternatively performing the following two steps:  

\begin{itemize}
  \item \textbf{Step-1: Solve (P11)} Given $\boldsymbol{\lambda}^{(t-1)}$ and $\boldsymbol{\mu}^{(t-1)}$, we solve (P10) and denote the solution by  $\big(\boldsymbol{\phi}^{(t)},\boldsymbol{V}^{(t)},\boldsymbol{\varpi}^{(t)}\big)$.
  
  \item \textbf{Step-2: Update $\boldsymbol{\lambda}$ and $\boldsymbol{\mu}$} Given $\big(\boldsymbol{\phi}^{(t)},\boldsymbol{V}^{(t)},\boldsymbol{\varpi}^{(t)}\big)$, we utilize the modified Newton (MN) method of~\cite{Jonga-Sum-of-Ratios} to obtain $\boldsymbol{\lambda}^{(t)}$ and $\boldsymbol{\mu}^{(t)}$. 
\end{itemize}

For the update in \textbf{Step-1}, (P11) is nonconvex in $(\boldsymbol{\phi},\boldsymbol{V},\boldsymbol{\varpi})$ jointly. Therefore we still adopt the BCD method to solve it. Recall the discussion explaining (\ref{wmmse_transform}), the optimal $\varpi$ is obtained in (\ref{w_opt}) when  $(\boldsymbol{\phi},\boldsymbol{V})$ is fixed, and the optimal $\boldsymbol{V}$ can be determined in (\ref{v_opt}) when $(\boldsymbol{\phi},\boldsymbol{\varpi})$ is fixed. Therefore we need to investigate the optimization of $\boldsymbol{\phi}$. When $(\boldsymbol{V},\boldsymbol{\varpi})$, the optimiztion of (P11) with respect to $\boldsymbol{\phi}$ can be written as

\begin{subequations}
\begin{alignat}{2}
\!\!\!\!\!\!\textup{(P12):} \max_{\boldsymbol{\phi}} \ & \boldsymbol{\phi}^H\boldsymbol{T}\boldsymbol{\phi}+2\mathrm{Re}\{\boldsymbol{t}^H\boldsymbol{\phi}\}+c\\
\sf{s.t.} &\ \boldsymbol{\phi}^H\boldsymbol{T}_k\boldsymbol{\phi}+2\mathrm{Re}\{\boldsymbol{t}_k^H\boldsymbol{\phi}\}+c_k\geq r_k, \forall k\in\mathcal{K}, \\
&\ \ |\boldsymbol{\phi}_n|\leq1, \forall n\in\mathcal{N}
\end{alignat}
\end{subequations}
with the parameters defined as follows:
\begin{subequations}
\begin{align}
\!\!\!\!\boldsymbol{T}_k\triangleq &-\varpi_k\Big[\sum_{j}q_j\boldsymbol{F}_j^H\boldsymbol{v}_k\boldsymbol{v}_k^H\boldsymbol{F}_j\Big], \label{eq_cnstr_T}\\ 
\boldsymbol{t}_k\triangleq & -\!\varpi_k\Big[q_k\big(\boldsymbol{v}_k^H\boldsymbol{h}_{{\rm{d}},k}\!-\!1\big)\boldsymbol{F}_k^H\boldsymbol{v}_k\!+\!\sum_{j\neq k}q_j\boldsymbol{v}_k^H\boldsymbol{h}_{{\rm{d}},j}\boldsymbol{F}_j^H\boldsymbol{v}_k\Big] \nonumber\\
c_k \triangleq & \log{\varpi_k}-\varpi_kq_k\!+\!1\!+\!\log{q_k}-\sigma^2\varpi_k\|\boldsymbol{v}_k\|_2^2 \\
&\quad -\varpi_k\sum_{j }q_j\big|\boldsymbol{h}_{{\rm{d}},j}^H\boldsymbol{v}_k\big|^2+2\varpi_kq_k\mathrm{Re}\big\{\boldsymbol{h}_{{\rm{d}},k}^H\boldsymbol{v}_k\big\} \nonumber \\
\boldsymbol{T} \triangleq & \sum_{k}\lambda_k\mu_k\boldsymbol{T}_k, 
\boldsymbol{t} \triangleq \sum_{k}\lambda_k\mu_k\boldsymbol{t}_k, 
c \triangleq \sum_{k}\lambda_k\mu_kc_k.
\end{align}
\end{subequations}
The above problem is convex and can be conveniently solved numerically. 

For the update in \textbf{Step-2}, we introduce the following notations to simplify the subsequent expositions. According to (\ref{lem-part1-lambda}) and~(\ref{lem-part1-gamma}) define
\begin{align}
\Lambda_k^{(t)}(\lambda_k) & \triangleq \lambda_k \widetilde{R}_k(\boldsymbol{\phi}^{(t)},\boldsymbol{v}_k^{(t)},\varpi_k^{(t)}) - 1,  \forall k\in\mathcal{K}, \label{eq-Lambda-k-t} \\
\Gamma_k^{(t)}(\mu_k) & \triangleq \mu_k\widetilde{R}_k(\boldsymbol{\phi}^{(t)},\boldsymbol{v}_k^{(t)},\varpi_k^{(t)}) - A_k,  \forall k\in\mathcal{K}. \label{eq-Gamma-k-t}
\end{align}
Then according to the modified Newton's (MN) method in \cite{Jonga-Sum-of-Ratios}, we update $\boldsymbol{\lambda}$ and $\boldsymbol{\mu}$ in the  following way
\begin{align}
\lambda_{k,i}^{(t)} & : = \lambda_k^{(t-1)} - \xi^i \frac{\Lambda_k^{(t)}(\lambda_k^{(t-1)})}{ \frac{\textup{d} \Lambda_k^{(t)}(\lambda_k) }{\textup{d}\lambda_k }\big|_{\lambda_k = \lambda_k^{(t-1)}} }\nonumber\\
&= \lambda_k^{(t-1)} - \xi^i \frac{\Lambda_k^{(t)}(\lambda_k^{(t-1)})}{\widetilde{R}_k(\boldsymbol{\phi}^{(t)},\boldsymbol{v}_k^{(t)},\varpi_k^{(t)})}, \forall k\in\mathcal{K} \label{def-lambda-k-i} \\
\mu_{k,i}^{(t)} & := \mu_k^{(t-1)} -\xi^i \frac{\Gamma_k^{(t)}(\mu_k^{(t-1)})}{ \frac{\textup{d} \Gamma_k^{(t)}(\mu_k) }{\textup{d}\mu_k }\big|_{\mu_k=\mu_k^{(t-1)}}} \nonumber \\
&= \mu_k^{(t-1)} -\xi^i  \frac{\Gamma_k^{(t)}(\mu_k^{(t-1)})}{\widetilde{R}_k(\boldsymbol{\phi}^{(t)},\boldsymbol{v}_k^{(t)},\varpi_k^{(t)})},\forall k\in\mathcal{K}. \label{def-gamma-k-i}
\end{align}

Clearly, (\ref{lem-part1-lambda}) and (\ref{lem-part1-gamma}) will reduce to the standard Newton's method update when $\xi^i$ above is $1$. In the modified Newton's method, a suitable $i$ should be found. Let $i^{(t)}$ denote the smallest integer among $i\in\{0,1,\ldots\}$ satisfying that
\begin{align}
&\sum_{k=1}^{K}  \Bigg\{\left(\Lambda_k^{(t)}( \lambda_{k,i}^{(t)}) \right)^2 + \sum_{k=1}^{K}  \left(\Gamma_k^{(t)}( \mu_{k,i}^{(t)}) \right)^2\Bigg\} \label{eq-update-rule} \\ 
&\leq \left(1-\xi^i\epsilon \right)^2  \left[\sum_{k=1}^{K}  \left(\Lambda_k^{(t)}( \lambda_{k}^{(t-1)}) \right)^2 + \sum_{k=1}^{K}  \left(\Gamma_k^{(t)}( \mu_{k}^{(t-1)}) \right)^2 \right], \nonumber 
\end{align}
where $\epsilon$ is a predefined precision parameter and $\xi$ is a positive constant smaller than $1$. Once $i^{(t)}$ is determined, $\boldsymbol{\lambda}^{(t)}$ and $\boldsymbol{\mu}^{(t)}$ should be updated as  
\begin{align}
\lambda_{k}^{(t)} & \leftarrow \lambda_{k,i^{(t)}}^{(t)} \textup{in  (\ref{def-lambda-k-i})}\textup{\ with $i$ being $i^{(t)}$} , \label{set-lambda-k-t-plus-1} \\
\mu_{k}^{(t)}& \leftarrow \mu_{k,i^{(t)}}^{(t)}  \textup{in (\ref{def-gamma-k-i})}\ \textup{with $i$ being $i^{(t)}$}, \label{set-gamma-k-t-plus-1}
\end{align}

According to the analysis in \cite{Jonga-Sum-of-Ratios} the value 
\begin{align}
\delta^{(t)}\triangleq\sum_{k=1}^{K}  \left(\Lambda_k^{(t)}( \lambda_{k}^{(t)}) \right)^2 + \sum_{k=1}^{K}  \left(\Gamma_k^{(t)}( \gamma_{k}^{(t)}) \right)^2 \label{eq-two-sum}
\end{align}
converges to zero as $t\rightarrow\infty$, and the rate of convergence is global linear and local quadratic. When $\delta^{(t)}$ in (\ref{eq-two-sum}) converges to zero, according to (\ref{eq-Lambda-k-t}) and (\ref{eq-Gamma-k-t})
\begin{align}
&\Lambda_k^{(t)}( \lambda_{k}^{(t)})\!=\!0 \Rightarrow  \lambda_k^{(t)} \widetilde{R}_k(\boldsymbol{\phi}^{(t)},\boldsymbol{v}_k^{(t)},\varpi_k^{(t)})\!\!=\!\!1,  \forall k\in\mathcal{K}, \label{eq-Lambda-k-t-terminate} \\
&\Gamma_k^{(t)}( \mu_{k}^{(t)})\!=\!0 \Rightarrow \mu_k^{(t)} \widetilde{R}_k(\boldsymbol{\phi}^{(t)},\boldsymbol{v}_k^{(t)},\varpi_k^{(t)})\!\!=\!\!A_k,  \forall k\in\mathcal{K}. \label{eq-Gamma-k-t-terminate}
\end{align}
Hence we will arrive at (\ref{lem-part1-lambda}) and (\ref{lem-part1-gamma}) and  consequently, by Lemmas~\ref{lem-part1} and~\ref{lem-part2}, the KKT condition of (P10). Therefore (\ref{eq-two-sum}), i.e. (\ref{eq-Lambda-k-t-terminate}) and (\ref{eq-Gamma-k-t-terminate}), performs as a termination condition for the iterative process.


\begin{algorithm}[!t]
\caption{Optimizing (P10)} \label{Alg-opt-overall}
\begin{algorithmic}[1]
\STATE Set $\xi\in(0,1)$; set the positive $\epsilon$, $\rho$ sufficiently small;
\STATE Invoke Alg.~\ref{Alg-feas-prob} to obtain a feasible $\boldsymbol{\phi}^{(0)}$; $t=0$;
\STATE Initialize $\boldsymbol{\varpi}^{(0)}$ and $\boldsymbol{V}^{(0)}$ via (\ref{w_opt}) and (\ref{v_opt}) respectively;
\STATE Set $\lambda_{k}^{(0)}=R_k(\boldsymbol{\phi}^{(0)})^{\!-\!1}$ and $\mu_{k}^{(0)}\!=\!A_k/R_k(\boldsymbol{\phi}^{(0)})$;
\REPEAT 
\REPEAT
\STATE update $\boldsymbol{\varpi}^{(t)}$ by (\ref{w_opt});
\STATE update $\boldsymbol{V}^{(t)}$ by (\ref{v_opt});
\STATE update $\boldsymbol{\phi}^{(t)}$ by solving (P12);
\UNTIL{\textit{convergence}}
\STATE update $\lambda_k^{(t\!+\!1)}$ and $\mu_k^{(t\!+\!1)}$ by (\ref{set-lambda-k-t-plus-1}) and (\ref{set-gamma-k-t-plus-1}) respectively;
\STATE Evaluate $\delta^{(t)}$ by (\ref{eq-update-rule});
\STATE  $t\leftarrow t+1$; 
\UNTIL{$\delta^{(0)}<\rho$}
\end{algorithmic}
\end{algorithm}

\section{Simulation} \label{sec-Simulation}

\begin{figure}[ht]
\centering
  \includegraphics[width=0.5\textwidth]{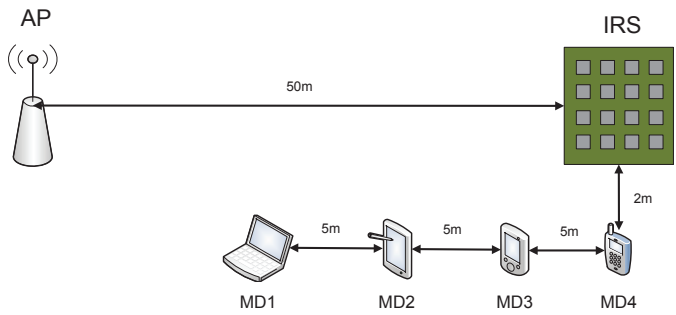} \vspace{-2pt}\caption{Simulation Setting-Up}
 \label{fig:system}
\end{figure}

In this section, we present numerical results to validate our proposed algorithms. In the experiment, we adopt a similar system setting-up reported in \cite{wu2018intelligent}. As shown in Fig.~\ref{fig:system}, the system comprises one AP with $M=4$ antennas, 4 single-antenna mobile users and one IRS device. The number of reflecting elements $N$ of the IRS will take different values from $\{30,60,90,120\}$. The distance between the AP and IRS is $50$m and it is assumed that the signal propagation environment between the AP and the IRS is dominated by the line-of-sight (LoS) link. Considering the mobility of the MDs will deteriorate the wireless propagation environment, we assume that the signals sent by MDs to both the AP and the IRS experience $10$dB penetration loss, independent Rayleigh fading and the pathloss exponent of $3$. The noise has -$170$dBm/Hz and the channel bandwidth is $100$KHz, so $\sigma_k^2 = 10^{-12}$mW. We set the transmission power of each MD as $10$dBm. Besides we assumed the antenna gain of both the AP and user is $0$dBi and that of each reflecting element at the IRS is $5$ dBi \cite{antennaRFID2009}. The distance between the AP and the  IRS is $50$m and the 4 MDs are located in a row parallel to the AP-IRS line and with a $5$m interval between the adjacent peers.  

\begin{figure}[ht]
\centering
  \includegraphics[width=0.5\textwidth]{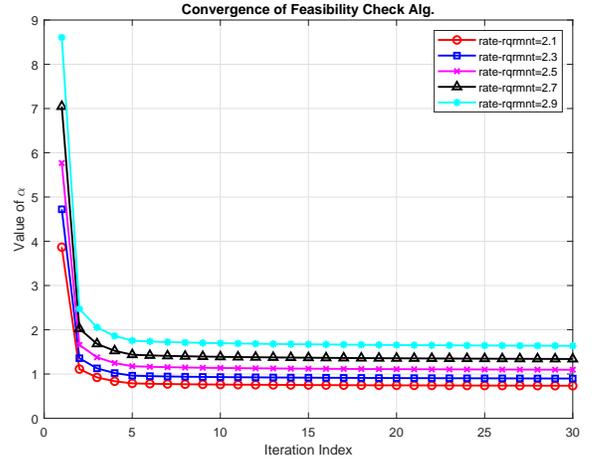} \vspace{-2pt}\caption{Convergence of Feasibility Check Alg.~\ref{Alg-feas-prob}}
 \label{fig:feas_check_convg}
\end{figure}

Fig.~\ref{fig:feas_check_convg} shows us the behavior of the Alg.~\ref{Alg-feas-prob} for checking the feasibility associated with a given set of information rate requirements. In our experiment, we set $N=30$. To ensure fairness between different MDs, we just assume that all MDs will be guaranteed one identical information rate. We vary this common information rate requirement from $2.1$Nats/s/channel-use to $2.9$Nats/s/channel-use and perform Alg.~\ref{Alg-feas-prob}. Note that in our experiment one random channel realization is generated according to system setting explained above and then fixed. For each specific information rate requirement, $\boldsymbol{\phi}$ is started from a common initial point, which is randomly chosen. According to Fig.~\ref{fig:feas_check_convg}, Alg.~\ref{Alg-feas-prob} generally converges within 10 iterations. The convergent $\alpha$ value increases when the information rate requirement inflates. It can be inferred from Fig.~\ref{fig:feas_check_convg} that maximal feasible common information rate requirement should lie between $2.3$ to $2.5$ Nats/s/channel-use.

\begin{figure}[ht]
\centering
  \includegraphics[width=0.5\textwidth]{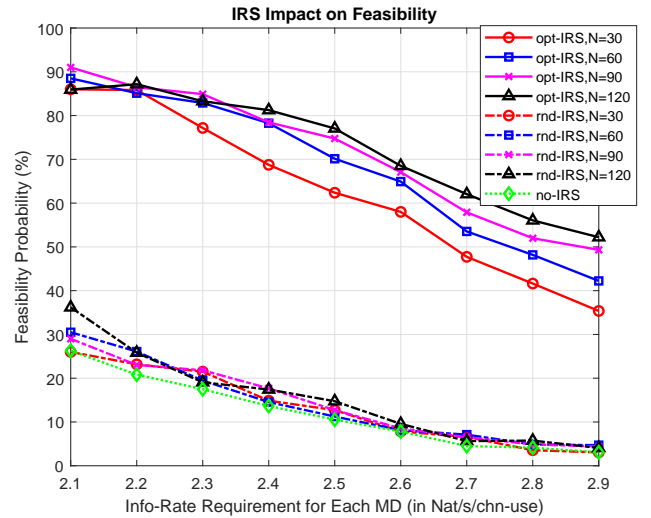} \vspace{-2pt}\caption{The Impact of IRS on Feasibility of Rate Requirements}
 \label{fig:irs_vs_feas}
\end{figure}

Fig.~\ref{fig:irs_vs_feas} illustrates the impact of IRS on enhancing the guaranteed information rate to MDs. In our experiment, we still assume that all MDs require one common information rate requirement. For each specific $N$ and a predefined rate requirement, $600$ channel realizations are randomly generated according to the above settings. For each channel realization, we generate a batch of random IRS elements and then invoke Alg.~\ref{Alg-feas-prob} to optimize the IRS elements trying to reach feasibility. We compare the probability of getting feasibility for the case of no-IRS, random IRS and optimized IRS in Fig.~\ref{fig:irs_vs_feas}. The result in Fig.~\ref{fig:irs_vs_feas} undoubtedly convince us the significance of IRS. Among the range of rate requirement in the figure, the optimized IRS can improve the feasibility probability by $40\%$ to $50\%$. Besides, more improvement could be obtained with more number of IRS elements being deployed.

\begin{figure}[ht]
\centering
  \includegraphics[width=0.5\textwidth]{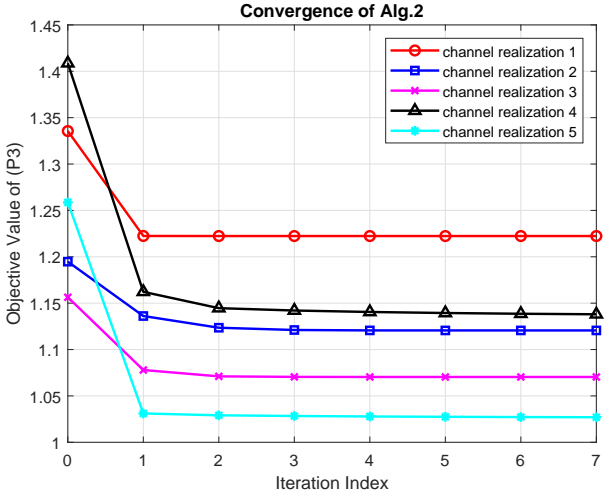} \vspace{-2pt}\caption{Convergence of Obj. of (P3) by Alg.~\ref{Alg-opt-overall}}
 \label{fig:itinery_sor_obj}
\end{figure}

\begin{figure}[ht]
\centering
  \includegraphics[width=0.5\textwidth]{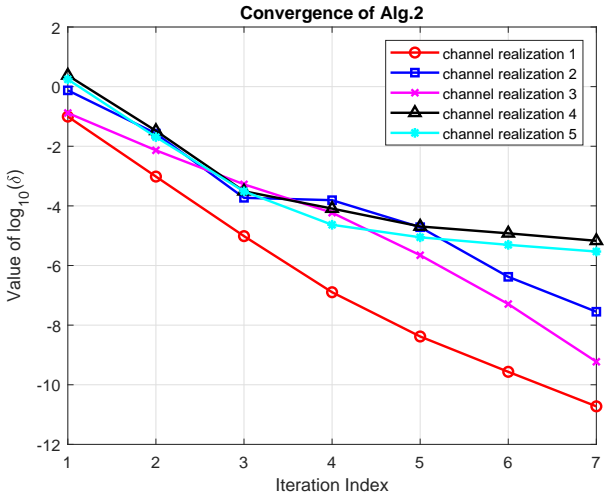} \vspace{-2pt}\caption{Convergence of $\delta^{(t)}$ by Alg.~\ref{Alg-opt-overall}}
 \label{fig:itinery_sor_delta}
\end{figure}

Fig.~\ref{fig:itinery_sor_obj} and Fig.~\ref{fig:itinery_sor_delta} show the convergence behavior of Alg.~\ref{Alg-opt-overall}. In the experiment we generate a group (5 in the figure) of random channel realizations, for which a feasible point can be found (e.g. by Alg.~\ref{Alg-feas-prob}). For each specific channel realization, we perform Alg.~\ref{Alg-opt-overall} starting from the feasible point. The objective value and the $\delta^{(t)}$ defined in (\ref{eq-two-sum}) are presented in Fig.~\ref{fig:itinery_sor_obj} and Fig.~\ref{fig:itinery_sor_delta} respectively. As shown in Fig.~\ref{fig:itinery_sor_obj}, Alg.~\ref{Alg-opt-overall} usually converges in several outer loops. Fig.~\ref{fig:itinery_sor_delta} convince us that Alg.~\ref{Alg-opt-overall} exhibits superlinear convergence rate. The results in Fig.~\ref{fig:itinery_sor_obj} and Fig.~\ref{fig:itinery_sor_delta} suggest that Alg.~\ref{Alg-opt-overall} has fast converge and is therefore very efficient.

\begin{figure}[ht]
\centering
  \includegraphics[width=0.5\textwidth]{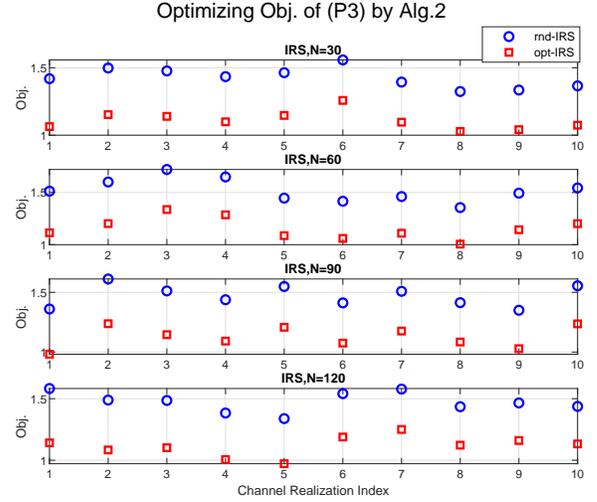} \vspace{-2pt}\caption{Optimizing (P3)}
 \label{fig:irs_vs_obj}
\end{figure}

Finally Fig.~\ref{fig:irs_vs_obj} illustrates the effect of Alg.~\ref{Alg-opt-overall} in optimizing (P3). Taking into account of fairness of different MDs, we just assume that $A_k$ are equal to each MDs and consequently set $A_k=1$ $\forall k\in \mathcal{K}$. For each specific $N$, we generate a bunch (10 in the figure) of random channel realizations and their feasible solutions which are found via randomization. We compare the objective values of (P3) obtained via Alg.~\ref{Alg-opt-overall} with those associated with those random IRS elements. As presented in Fig.~\ref{fig:irs_vs_obj}, for various values of $N$, Alg.~\ref{Alg-opt-overall} can effectively decrease the objective value of (P3) and therefore equivalently increase the earnings of the edge server.

\section{Conclusion} \label{sec-Conclusion}

In this paper we study the mobile edge computing problem with the aid of the novel technology of IRS. In details, we formulate the earning maximization problem with a group of information rate constraints in the presence of IRS device. We develop an iterative algorithm which can efficiently identify the feasibility of the information rate requirements and find a feasible solution. Besides we also develop an algorithm to optimize the earning of the edge server for loading computing. Substantial numerical results have shown that our proposed IRS aided scheme can significantly improve the guaranteed information rate to MDs by the AP and also effectively enlarge the earning of the edge server.

\appendix

\subsection{Proof of Lemma~\ref{lem-part1}} 
\label{subsec-proof_lem1}
\begin{proof}
To see the statements hold, we first give out the KKT conditions of (P10). To simplify the following expositions, we define
$\boldsymbol{x}_k\triangleq\big(\boldsymbol{\phi},\boldsymbol{v}_k,\varpi_k\big)$. Then the Lagrangian function is given as   
\begin{align}
\mathcal{L}&\big(\{\boldsymbol{x}_k\},\boldsymbol{\mu},\boldsymbol{\lambda},\boldsymbol{\zeta},\boldsymbol{\nu}\big) \triangleq\sum_k\mu_k+\sum_k\lambda_k\Big(A_k\!-\!\mu_k\widetilde{R}_k(\boldsymbol{x}_k)\Big)\nonumber \\
&\qquad+\sum_{k}\zeta_k\Big(r_k-\widetilde{R}_k(\boldsymbol{x}_k)\Big)+\sum_k\nu_k\Big(|\boldsymbol{\phi}_n|-1\Big),
\end{align}
with parameters $\{\lambda_k\}$, $\{\zeta_k\}$ and $\{\nu_k\}$ being the Lagrangian multipliers associated with the constraints (\ref{p11_frac_cnstr}), (\ref{p11_rate_cnstr}) and (\ref{phi_cnstr}) respectively. The KKT conditions are listed as follows:
\begin{subequations}
\label{KKT_cond}
\begin{align}
&\frac{\partial\mathcal{L}}{\partial\boldsymbol{x}_k^*}=\boldsymbol{0}, \forall k \in \mathcal{K},\label{kkt_x}\\
&\frac{\partial\mathcal{L}}{\partial\mu_k}=1-\lambda_k\widetilde{R}_k\big(\boldsymbol{x}_k\big)=0, \forall k \in \mathcal{K},\label{kkt_mu}\\
&\frac{\partial\mathcal{L}}{\partial\lambda_k}=A_k-\mu_k\widetilde{R}_k\big(\boldsymbol{x}_k\big)=0, \forall k \in\mathcal{K},\label{kkt_lambda}\\
&\frac{\partial\mathcal{L}}{\partial\zeta_k}=0, \forall k \in \mathcal{K}, \label{kkt_zeta}\\
&\frac{\partial\mathcal{L}}{\partial\nu_k}=0, \forall k \in \mathcal{K}, \label{kkt_nu}\\
&\quad\lambda_k\geq0, \zeta_k\geq0, \nu_k\geq0, \forall k\in\mathcal{K}\label{kkt_inequ}.
\end{align}
\end{subequations}
Assume that $(\boldsymbol{\phi}^\star,\boldsymbol{V}^\star,\boldsymbol{\varpi}^\star,\boldsymbol{\mu}^\star)$ is an optimal solution to (P10). Then there exist Lagrangian multipliers $\boldsymbol{\lambda}^\star$, $\boldsymbol{\zeta}^\star$ and $\boldsymbol{\nu}^\star$ such that the KKT conditions are satisfied. By (\ref{kkt_mu}) and (\ref{kkt_lambda}) we obtain the results in (\ref{lem-part1-lambda}) and (\ref{lem-part1-gamma}). Besides, it can be readily seen that the KKT conditions (\ref{kkt_x}), (\ref{kkt_zeta}), (\ref{kkt_nu}) and (\ref{kkt_inequ}) with $\boldsymbol{x}^\star$,$\boldsymbol{\zeta}^\star$ and $\boldsymbol{\nu}^\star$ being substituted therein are just the KKT conditions of (P11) with its parameters setting as $\boldsymbol{\lambda}=\boldsymbol{\lambda}^\star$ and $\boldsymbol{\mu}=\boldsymbol{\mu}^\star$. 
\end{proof}

\subsection{Proof of Lemma~\ref{lem-part2}} 
\label{subsec-proof_lem2}
\begin{proof}
Here we follow the notations used in the proof of Lemma \ref{lem-part1}. Assume that $(\boldsymbol{\phi}^\star,\boldsymbol{V}^\star,\boldsymbol{\varpi}^\star)$ solves the problem (P11) when its parameters are set as $\boldsymbol{\lambda}=\boldsymbol{\lambda}^\star$ and $\boldsymbol{\mu}=\boldsymbol{\mu}^\star$. Then by optimality conditions, the KKT conditions of (P11) should be satisfied with by $(\boldsymbol{\phi}^\star,\boldsymbol{V}^\star,\boldsymbol{\varpi}^\star)$ together with the Lagrangian multipliers $\boldsymbol{\zeta}^\star$ and $\boldsymbol{\nu}^\star$ associated with the constraints (\ref{reduced_func_rate_cnstr}) and (\ref{reduced_func_model_cnstr}) respectively. That is the equations (\ref{kkt_x}), (\ref{kkt_zeta}), (\ref{kkt_nu}) and (\ref{kkt_inequ}) hold. Taking into account the conditions in \ding{173}, the equations (\ref{kkt_mu}) and (\ref{kkt_lambda}) also stand. Therefore (\ref{KKT_cond}) are satisfied, which are indeed the KKT conditions of (P10) associated with the optimal solution $(\boldsymbol{\phi}^\star,\boldsymbol{V}^\star,\boldsymbol{\varpi}^\star,\boldsymbol{\mu}^\star)$. Therefore the proof is complete.
\end{proof}











\bibliographystyle{IEEEtran}
\bibliography{related}

\begin{thebibliography}{10}
\providecommand{\url}[1]{#1}
\csname url@samestyle\endcsname
\providecommand{\newblock}{\relax}
\providecommand{\bibinfo}[2]{#2}
\providecommand{\BIBentrySTDinterwordspacing}{\spaceskip=0pt\relax}
\providecommand{\BIBentryALTinterwordstretchfactor}{4}
\providecommand{\BIBentryALTinterwordspacing}{\spaceskip=\fontdimen2\font plus
\BIBentryALTinterwordstretchfactor\fontdimen3\font minus
  \fontdimen4\font\relax}
\providecommand{\BIBforeignlanguage}[2]{{%
\expandafter\ifx\csname l@#1\endcsname\relax
\typeout{** WARNING: IEEEtran.bst: No hyphenation pattern has been}%
\typeout{** loaded for the language `#1'. Using the pattern for}%
\typeout{** the default language instead.}%
\else
\language=\csname l@#1\endcsname
\fi
#2}}
\providecommand{\BIBdecl}{\relax}
\BIBdecl

\bibitem{satyanarayanan2017emergence}
M.~Satyanarayanan, ``The emergence of edge computing,'' \emph{Computer},
  vol.~50, no.~1, pp. 30--39, 2017.

\bibitem{mao2017survey}
Y.~Mao, C.~You, J.~Zhang, K.~Huang, and K.~B. Letaief, ``A survey on mobile
  edge computing: {The} communication perspective,'' \emph{IEEE Communications
  Surveys \& Tutorials}, vol.~19, no.~4, pp. 2322--2358, 2017.

\bibitem{hu2018wireless}
X.~Hu, K.-K. Wong, and K.~Yang, ``Wireless powered cooperation-assisted mobile
  edge computing,'' \emph{IEEE Transactions on Wireless Communications},
  vol.~17, no.~4, pp. 2375--2388, 2018.

\bibitem{vaughan2019working}
O.~Vaughan, ``Working on the edge,'' \emph{Nature Electronics}, vol.~2, no.~1,
  p.~2, 2019.

\bibitem{Microsoft}
``{AT\&T} and {Microsoft} announce a strategic alliance to deliver innovation
  with cloud, {AI} and {5G},''
  \url{https://techblog.comsoc.org/2019/07/17/att-announced-cloud-partnerships-with-microsoft-1-day-after-similar-deal-with-ibm/},
  accessed: 2019-07-19.

\bibitem{mach2017mobile}
P.~Mach and Z.~Becvar, ``Mobile edge computing: {A} survey on architecture and
  computation offloading,'' \emph{IEEE Communications Surveys \& Tutorials},
  vol.~19, no.~3, pp. 1628--1656, 2017.

\bibitem{chen2014decentralized}
X.~Chen, ``Decentralized computation offloading game for mobile cloud
  computing,'' \emph{IEEE Transactions on Parallel and Distributed Systems},
  vol.~26, no.~4, pp. 974--983, 2014.

\bibitem{chen2015efficient}
X.~Chen, L.~Jiao, W.~Li, and X.~Fu, ``Efficient multi-user computation
  offloading for mobile-edge cloud computing,'' \emph{IEEE/ACM Transactions on
  Networking}, vol.~24, no.~5, pp. 2795--2808, 2015.

\bibitem{tao2017performance}
X.~Tao, K.~Ota, M.~Dong, H.~Qi, and K.~Li, ``Performance guaranteed computation
  offloading for mobile-edge cloud computing,'' \emph{IEEE Wireless
  Communications Letters}, vol.~6, no.~6, pp. 774--777, 2017.

\bibitem{mao2016dynamic}
Y.~Mao, J.~Zhang, and K.~B. Letaief, ``Dynamic computation offloading for
  mobile-edge computing with energy harvesting devices,'' \emph{IEEE Journal on
  Selected Areas in Communications}, vol.~34, no.~12, pp. 3590--3605, 2016.

\bibitem{DOCOMO}
``{NTT DoCoMo} and {Metawave} announce successful demonstration of {28GHz}-band
  {5G} using world's first meta-structure technology,''
  \url{https://www.marketwatch.com/press-release/ntt-docomo-and-metawave-announce-successful-demonstration-of-28ghz-band-5g-using-worlds-first-meta-structure-technology-2018-12-04},
  accessed: July 7, 2019.

\bibitem{hu2018beyond}
S.~Hu, F.~Rusek, and O.~Edfors, ``Beyond massive {{MIMO}}: {The} potential of
  data transmission with large intelligent surfaces,'' \emph{IEEE Transactions
  on Signal Processing}, vol.~66, no.~10, pp. 2746--2758, 2018.

\bibitem{CunhuaPan_MEC_IRS}
T.~Bai, C.~Pan, Y.~Deng, M.~Elkashlan, and A.~Nallanathan, ``Latency
  minimization for intelligent reflecting surface aided mobile edge
  computing,'' \emph{arXiv preprint arXiv:1910.07990v1}, 2019.

\bibitem{hu2015mobile}
Y.~C. Hu, M.~Patel, D.~Sabella, N.~Sprecher, and V.~Young, ``Mobile edge
  computing—a key technology towards {5G},'' \emph{ETSI white paper},
  vol.~11, no.~11, pp. 1--16, 2015.

\bibitem{bonomi2012fog}
F.~Bonomi, R.~Milito, J.~Zhu, and S.~Addepalli, ``Fog computing and its role in
  the internet of things,'' in \emph{ACM Mobile Cloud Computing (MCC)}, 2012,
  pp. 13--16.

\bibitem{IEEEOpenFog}
``{IEEE} standard for adoption of {OpenFog} reference architecture for fog
  computing,'' \url{https://standards.ieee.org/standard/1934-2018.html},
  accessed: 2019-07-19.

\bibitem{wang2019edge}
J.~Wang, J.~Pan, F.~Esposito, P.~Calyam, Z.~Yang, and P.~Mohapatra, ``Edge
  cloud offloading algorithms: {Issues}, methods, and perspectives,'' \emph{ACM
  Computing Surveys (CSUR)}, vol.~52, no.~1, p.~2, 2019.

\bibitem{wang2016mobile}
Y.~Wang, M.~Sheng, X.~Wang, L.~Wang, and J.~Li, ``Mobile-edge computing:
  {Partial} computation offloading using dynamic voltage scaling,'' \emph{IEEE
  Transactions on Communications}, vol.~64, no.~10, pp. 4268--4282, 2016.

\bibitem{ning2018cooperative}
Z.~Ning, P.~Dong, X.~Kong, and F.~Xia, ``A cooperative partial computation
  offloading scheme for mobile edge computing enabled {Internet of Things},''
  \emph{IEEE Internet of Things Journal}, 2018.

\bibitem{bi2018computation}
S.~Bi and Y.~J. Zhang, ``Computation rate maximization for wireless powered
  mobile-edge computing with binary computation offloading,'' \emph{IEEE
  Transactions on Wireless Communications}, vol.~17, no.~6, pp. 4177--4190,
  2018.

\bibitem{chen2018task}
M.~Chen and Y.~Hao, ``Task offloading for mobile edge computing in software
  defined ultra-dense network,'' \emph{IEEE Journal on Selected Areas in
  Communications}, vol.~36, no.~3, pp. 587--597, 2018.

\bibitem{Junginproceedings}
\BIBentryALTinterwordspacing
M.~Jung, W.~Saad, Y.~Jang, G.~Kong, and S.~Choi, ``Uplink data rate in large
  intelligent surfaces: {Asymptotic} analysis under channel estimation
  errors,'' 2018. [Online]. Available:
  \url{https://www.researchgate.net/publication/328827179_Uplink_Data_Rate_in_Large_Intelligent_Surfaces_Asymptotic_Analysis_under_Channel_Estimation_Errors}
\BIBentrySTDinterwordspacing

\bibitem{jung2018performance}
------, ``Performance analysis of large intelligence surfaces {(LISs)}:
  {Asymptotic} data rate and channel hardening effects,'' \emph{arXiv preprint
  arXiv:1810.05667}, 2018.

\bibitem{huang2019Reconfigurable}
C.~Huang, A.~Zappone, G.~C. Alexandropoulos, M.~Debbah, and C.~Yuen,
  ``Reconfigurable intelligent surfaces for energy efficiency in wireless
  communication,'' \emph{IEEE Transactions on Wireless Communications}, 2019.

\bibitem{fu2019intelligent}
M.~Fu, Y.~Zhou, and Y.~Shi, ``Intelligent reflecting surface for downlink
  non-orthogonal multiple access networks,'' \emph{arXiv preprint
  arXiv:1906.09434}, 2019.

\bibitem{yu2019miso}
X.~Yu, D.~Xu, and R.~Schober, ``{MISO} wireless communication systems via
  intelligent reflecting surfaces,'' \emph{arXiv preprint arXiv:1904.12199},
  2019.

\bibitem{nadeem2019intelligent}
Q.-U.-A. Nadeem, A.~Kammoun, A.~Chaaban, M.~Debbah, and M.-S. Alouini,
  ``Intelligent reflecting surface assisted multi-user {MISO} communication,''
  \emph{arXiv preprint arXiv:1906.02360}, 2019.

\bibitem{mishra2019channel}
D.~Mishra and H.~Johansson, ``Channel estimation and low-complexity beamforming
  design for passive intelligent surface assisted {MISO} wireless energy
  transfer,'' in \emph{IEEE International Conference on Acoustics, Speech and
  Signal Processing (ICASSP)}, 2019, pp. 4659--4663.

\bibitem{jiang2019over}
T.~Jiang and Y.~Shi, ``Over-the-air computation via intelligent reflecting
  surfaces,'' \emph{arXiv preprint arXiv:1904.12475}, 2019.

\bibitem{qingqing2019towards}
W.~Qingqing and Z.~Rui, ``Towards smart and reconfigurable environment:
  {Intelligent} reflecting surface aided wireless network,'' \emph{arXiv
  preprint arXiv:1905.00152}, 2019.

\bibitem{basar2019large}
E.~Basar, ``Large intelligent surface-based index modulation: A new beyond
  {MIMO} paradigm for {6G},'' \emph{arXiv preprint arXiv:1904.06704}, 2019.

\bibitem{huang2018achievable}
C.~Huang, A.~Zappone, M.~Debbah, and C.~Yuen, ``Achievable rate maximization by
  passive intelligent mirrors,'' in \emph{IEEE International Conference on
  Acoustics, Speech and Signal Processing (ICASSP)}, 2018, pp. 3714--3718.

\bibitem{di2019reflection}
M.~Di~Renzo and J.~Song, ``Reflection probability in wireless networks with
  metasurface-coated environmental objects: {An} approach based on random
  spatial processes,'' \emph{arXiv preprint arXiv:1901.01046}, 2019.

\bibitem{nadeem2019large}
Q.-U.-A. Nadeem, A.~Kammoun, A.~Chaaban, M.~Debbah, and M.-S. Alouini,
  ``Asymptotic analysis of large intelligent surface assisted {MIMO}
  communication,'' \emph{arXiv preprint arXiv:1903.08127}, 2019.

\bibitem{wu2018intelligent}
Q.~Wu and R.~Zhang, ``Intelligent reflecting surface enhanced wireless network:
  Joint active and passive beamforming design,'' in \emph{IEEE Global
  Communications Conference (GLOBECOM)}, 2018, pp. 1--6.

\bibitem{wu2018intelligentfull}
------, ``Intelligent reflecting surface enhanced wireless network via joint
  active and passive beamforming,'' \emph{arXiv preprint arXiv:1810.03961},
  2018.

\bibitem{jung2019performance}
M.~Jung, W.~Saad, and G.~Kong, ``Performance analysis of large intelligent
  surfaces {(LISs)}: {Uplink} spectral efficiency and pilot training,''
  \emph{arXiv preprint arXiv:1904.00453}, 2019.

\bibitem{hu2017potential}
S.~Hu, F.~Rusek, and O.~Edfors, ``The potential of using large antenna arrays
  on intelligent surfaces,'' in \emph{IEEE 85th Vehicular Technology Conference
  (VTC Spring)}, 2017, pp. 1--6.

\bibitem{hu2018capacity}
------, ``Capacity degradation with modeling hardware impairment in large
  intelligent surface,'' in \emph{IEEE Global Communications Conference
  (GLOBECOM)}, 2018, pp. 1--6.

\bibitem{schubert2005iterative}
M.~Schubert and H.~Boche, ``Iterative multiuser uplink and downlink beamforming
  under {SINR} constraints,'' \emph{IEEE Transactions on Signal Processing},
  vol.~53, no.~7, pp. 2324--2334, 2005.

\bibitem{monzingo1980introduction}
R.~A. Monzingo and T.~W. Miller, \emph{Introduction to Adaptive Arrays}.\hskip
  1em plus 0.5em minus 0.4em\relax New York, Wiley-Interscience, 1980.

\bibitem{shannon1948mathematical}
C.~E. Shannon, ``A mathematical theory of communication,'' \emph{Bell System
  Technical Journal}, vol.~27, no.~3, pp. 379--423, 1948.

\bibitem{Cioffi_wmmse}
S.~Christensen, R.~Argawal, E.~de~Carvalho, and J.~M. Cioffi, ``Weighted
  sum-rate maximization using weighted mmse for mimo-bc beamforming design,''
  \emph{IEEE Transactions on Wireless Communication}, vol.~7, no.~12, pp. 1--7,
  2008.

\bibitem{QingjiangShi_wmmse}
Q.~Shi, M.~Razaviyayn, Z.-Q. Luo, and C.~Chen, ``An iteratively weighted mmse
  approach to distributed sum-utility maximization for a mimo interfering
  broadcast channel,'' \emph{IEEE Transactions on Signal Processing}, vol.~59,
  no.~9, pp. 4331--4340, 2011.

\bibitem{MEC_IRS_fullpaper}
Y.~Liu, J.~Zhao, Z.~Xiong, D.~Niyato, Y.~Chau, P.~Cunhua, and B.~Huang,
  ``Intelligent reflecting surface meets mobile edge computing: Enhancing
  wireless communications for computation offloading,'' 2020, full version of
  this \mbox{paper}. Available online at \\
  \url{http://www.ntu.edu.sg/home/junzhao/MEC_IRS.pdf}.

\bibitem{Jonga-Sum-of-Ratios}
Y.-C. Jong, ``An efficient global optimization algorithm for nonlinear
  sum-of-ratios problem,''
  \url{http://www.optimization-online.org/DB_FILE/2012/08/3586.pdf}, accessed:
  2020-01-10. A shorter version at arXiv:1207.1153.

\bibitem{antennaRFID2009}
J.~D. Griffin and G.~D. Durgin, ``Complete link budgets for backscatter-radio
  and rfid systems,'' \emph{IEEE Antennas Propagation Magazine}, vol.~51,
  no.~2, pp. 11--25, 2009.

\end{thebibliography}

\end{document}